\documentclass[11pt]{article}

\newsavebox{\foobox}
\newcommand{\slantbox}[2][0]{\mbox{%
        \sbox{\foobox}{#2}%
        \hskip\wd\foobox
        \pdfsave
        \pdfsetmatrix{1 0 #1 1}%
        \llap{\usebox{\foobox}}%
        \pdfrestore
}}
\newcommand\unslant[2][-.25]{\slantbox[#1]{$#2$}}

\newcommand{\mpi}{\text{\unslant[-.18]\pi}}
\newcommand{\mdelta}{\text{\unslant[-.18]\delta}}

\usepackage[left=2cm, right=2cm, top=2.5cm, bottom=2.5cm]{geometry}
\usepackage[x11names]{xcolor}
\usepackage{fancyhdr, amssymb, cancel, amsmath, graphicx, pgfplots, tikz}
\usepackage{isomath}

\usetikzlibrary{shadows}

\newcommand{\eff}{\operatorname{eff.}}

\renewcommand{\bar}{\overline}
\renewcommand{\tilde}{\widetilde}
\renewcommand{\hat}{\widehat}
\renewcommand{\le}{\leqslant}
\renewcommand{\ge}{\geqslant}
\renewcommand{\leq}{\leqslant}

\newcommand{\Pf}{\operatorname{Pf}}

\newcommand{\sgn}{\operatorname{sgn}}
\newcommand{\calJ}{J}
\newcommand{\calF}{\mathcal{F}}

\newcommand{\stylecolor}{IndianRed3}

\usepackage[labelfont={bf,sf, color=\stylecolor}, margin={1.5cm,0cm}]{caption}

\usepackage[colorlinks=true, urlcolor=\stylecolor, linkcolor=\stylecolor, citecolor=\stylecolor, hyperindex=true, linktocpage=true]{hyperref}

\usepackage{amsthm}
\newtheoremstyle{theor}{10pt}{10pt}{}{16pt}{\sffamily \bfseries \color{green!50!black}}{:}{.5em}{}
\theoremstyle{theor}

\usepackage[explicit]{titlesec}

\newcommand*\sectionlabel{}
\titleformat{\section}
  {\gdef\sectionlabel{}
   \Large\bfseries\scshape}
  {\gdef\sectionlabel{\thesection }}{0pt}
  {\begin{tikzpicture}[remember picture,overlay]
       \end{tikzpicture}
  }
\titlespacing*{\section}{0pt}{0pt}{0pt}

\newcommand*\subsectionlabel{}
\titleformat{\subsection}
  {\gdef\subsectionlabel{}
   \large\bfseries\scshape}
  {\gdef\subsectionlabel{\thesubsection  }}{0pt}
  {\begin{tikzpicture}[remember picture,overlay]
    	\draw (-0.15, 0.02) node[right] {\color{\stylecolor} \textsf{\subsectionlabel}};
	\draw (1.25, 0) node[right] {\color{\stylecolor} \textsf{#1}};
	\fill[color=\stylecolor] (1,-0.25) rectangle (1.1, 0.25);
       \end{tikzpicture}
  }
\titlespacing*{\subsection}{0pt}{10pt}{10pt}

\newcommand*\subsubsectionlabel{}
\titleformat{\subsubsection}
  {\gdef\subsubsectionlabel{}
   \bfseries\scshape}
  {\gdef\subsubsectionlabel{\thesubsubsection.\ \  }}{0pt}
  {\begin{tikzpicture}[remember picture,overlay]
    	\draw (-0.15, 0) node[right] {\color{\stylecolor} \textsf{\subsubsectionlabel#1}};
       \end{tikzpicture}
  }
\titlespacing*{\subsubsection}{0pt}{7pt}{7pt}

\pgfplotsset{every axis legend/.append style={at={(1.02,1)},anchor=north west}}

\begin{document}

\allowdisplaybreaks

\pagestyle{fancy}
\renewcommand{\headrulewidth}{0pt}
\fancyhead{}

\fancyfoot{}
\fancyfoot[C] {\textsf{\textbf{\thepage}}}

\begin{equation*}
\begin{tikzpicture}
\draw (\textwidth, 0) node[text width = \textwidth, right] {\color{white} easter egg};
\end{tikzpicture}
\end{equation*}

\begin{equation*}
\begin{tikzpicture}
\draw (0.5\textwidth, -3) node[text width = \textwidth] {\huge  \textsf{\textbf{Fast scrambling on sparse graphs}} };
\end{tikzpicture}
\end{equation*}
\begin{equation*}
\begin{tikzpicture}
\draw (0.5\textwidth, 0.1) node[text width=\textwidth] {\large \color{black} \textsf{Gregory Bentsen,}$^{\color{\stylecolor} \mathsf{a}}$ \textsf{Yingfei Gu}$^{\color{\stylecolor} \mathsf{b}}$  \textsf{and Andrew Lucas}$^{\color{\stylecolor} \mathsf{a}}$};
\draw (0.5\textwidth, -0.5) node[text width=\textwidth] {$^{\color{\stylecolor} \mathsf{a}}$ \small{\textsf{Department of Physics, Stanford University, Stanford, CA 94305, USA}}};
\draw (0.5\textwidth, -0.6) node[text width=\textwidth, below] {$^{\color{\stylecolor} \mathsf{b}}$ {\small \textsf{Department of Physics, Harvard University, Cambridge, MA 02138 USA}}};
\end{tikzpicture}
\end{equation*}
\begin{equation*}
\begin{tikzpicture}
\draw (0, -13.15) node[right, text width=0.5\paperwidth] {\texttt{ajlucas@stanford.edu}};
\draw (\textwidth, -13.1) node[left] {\textsf{April 2, 2019}};
\end{tikzpicture}
\end{equation*}
\begin{equation*}
\begin{tikzpicture}
\draw[very thick, color=\stylecolor] (0.0\textwidth, -5.75) -- (0.99\textwidth, -5.75);
\draw (0.12\textwidth, -6.25) node[left] {\color{\stylecolor}  \textsf{\textbf{Abstract:}}};
\draw (0.53\textwidth, -6) node[below, text width=0.8\textwidth, text justified] {\small Given a quantum many-body system with few-body interactions, how rapidly can quantum information be hidden during time evolution?  The fast scrambling conjecture is that the time to thoroughly mix information among $N$ degrees of freedom grows at least logarithmically in $N$.   We derive this inequality for generic quantum systems at infinite temperature,  bounding the scrambling time by a finite decay time of local quantum correlations at late times.  Using Lieb-Robinson bounds, generalized Sachdev-Ye-Kitaev models, and random unitary circuits, we propose that a logarithmic scrambling time  can be achieved in most quantum systems with sparse connectivity.  These models also elucidate how quantum chaos is not universally related to scrambling:  we construct random few-body circuits with  infinite Lyapunov exponent but logarithmic scrambling time.   We discuss analogies between quantum models on graphs and quantum black holes, and suggest methods to experimentally study scrambling with as many as 100 sparsely-connected quantum degrees of freedom.    };
\end{tikzpicture}
\end{equation*}

\tableofcontents

\titleformat{\section}
  {\gdef\sectionlabel{}
   \Large\bfseries\scshape}
  {\gdef\sectionlabel{\thesection }}{0pt}
  {\begin{tikzpicture}[remember picture]
	\draw (0.2, 0) node[right] {\color{\stylecolor} \textsf{#1}};
	\draw (0.0, 0) node[left, fill=\stylecolor,minimum height=0.27in, minimum width=0.27in] {\color{white} \textsf{\sectionlabel}};
       \end{tikzpicture}
  }
\titlespacing*{\section}{0pt}{20pt}{5pt}

\begin{equation*}
\begin{tikzpicture}
\draw[very thick, color=\stylecolor] (0.0\textwidth, -5.75) -- (0.99\textwidth, -5.75);
\end{tikzpicture}
\end{equation*}

\section{Introduction}
Despite  the unitarity of time evolution in quantum mechanics, it is still possible for an isolated many-body quantum system to thermalize and dissipate.  The origin of thermalization in such systems  is the spreading of quantum information and entanglement.  Information which was once stored in small regions of space, or in simple operators, can only be accessed by measuring a finite fraction of all $N\gg 1$ degrees of freedom at late times.   Our inability  to perform such complicated measurements is why these closed quantum systems  still appear thermal.

Understanding the process of thermalization in quantum systems is of great interest for a  number of reasons.  Firstly, a quantitative understanding of quantum thermalization, together with the emergence of  hydrodynamics, remains an incredibly hard problem \cite{polkovnikov} with experimental applications in nuclear, condensed matter and atomic physics.   Secondly, as quantum information has been hidden -- but not destroyed -- under unitary time evolution,  constraints on thermalization also imply bounds on quantum information processing and protection \cite{hayden07}.

A natural question is whether thermalization  can occur arbitrarily fast.  Remarkably, evidence from black  hole physics has led to the conjecture that fundamental bounds on the ``scrambling'' of quantum information do exist: the time $t_{\mathrm{s}}$ before quantum information can be thoroughly lost to a `local' observer obeys \cite{susskind08} \begin{equation}
t_{\mathrm{s}} \gtrsim  \frac{\log N}{\lambda_{\mathrm{s}}},  \;\;\;\; (N\rightarrow \infty) \label{eq:introbound}
\end{equation}   
with $\lambda_{\mathrm{s}}\lesssim N^0$.  This conjecture comes from a thought experiment where black holes appear to clone quantum states (in violation of a theorem \cite{zurek, dieks}) unless $t_{\mathrm{s}}$ obeys (\ref{eq:introbound}).  Three immediate questions arise.  Can this conjecture be understood on general quantum mechanical grounds, without appealing to black holes?  If so, what is $\lambda_{\mathrm{s}}$?   In which systems can the bound (\ref{eq:introbound}) be saturated?  The purpose of this paper is to explore and  resolve such questions in a broad class of quantum systems.

We first clarify the conjecture.  For simplicity, consider a  Hilbert space $\mathcal{H}$ which is a tensor product of $N$ two-dimensional Hilbert spaces $\mathcal{H}_i$.  We refer to each $\mathcal{H}_i$ as a single \emph{degree of freedom} (DOF).  We shall be concerned with the class of Hamiltonians $H$ acting on $\mathcal{H}$ that are $k$-local in the computer science sense:  namely, they involve  products of at most $k$ Pauli matrices $\sigma^\alpha_i$: \begin{align}
H &= \sum_{i=1}^N \sum_{\alpha=1}^3  h_i^\alpha \sigma_i^\alpha + \sum_{i,j=1}^N \sum_{\alpha,\beta=1}^3  h_{ij}^{\alpha \beta} \sigma_i^\alpha \sigma_j^\beta  +  \cdots \notag \\
&+ \sum_{i_1\cdots i_k, \alpha_1\cdots \alpha_k} h_{i_1\cdots i_k}^{\alpha_1 \cdots \alpha_k} \sigma_{i_1}^{\alpha_1}\cdots \sigma_{i_k}^{\alpha_k}.  \label{eq:introH}
\end{align} 
We will always take $k$ to be finite in the thermodynamic  limit $N\rightarrow  \infty$.   The coefficients above are normalized so that the Hamiltonian is extensive. 
We define the scrambling time as the time after which two initially unentangled halves of a system become nearly maximally entangled.   Let $A \subset \lbrace 1,\ldots, N \rbrace$ be a subset of $N^\prime=pN$
 of the degrees of freedom, with $0<p\le\frac{1}{2}$, such that $\mathcal{H} = \mathcal{H}_A \otimes \mathcal{H}_{A^c}$.   Given a time-dependent pure  state $|\Psi(t)\rangle$, we define the reduced density matrix of $A$:  \begin{equation}
\rho_A(t) = \underset{A^{\mathrm{c}}}{\mathrm{tr}} |\Psi(t)\rangle\langle \Psi(t)|,
\end{equation}
and the von Neumann entanglement entropy \begin{equation}
S_A(t) =  -\underset{A}{\mathrm{tr}}\left[\rho_A(t) \log \rho_A(t)\right].
\end{equation}
If the initial state $|\Psi(0)\rangle  = |\Psi_A\rangle \otimes |\Psi_{A^{\mathrm{c}}}\rangle$ has no entanglement between $A$ and its  complement, then  we define the scrambling time  as the smallest time when  \cite{susskind08}
  \begin{equation}
S_A(t_{\mathrm{s}}) \ge S_A^{\mathrm{max}} - a, \label{eq:SAmax}
\end{equation} regardless of the choice of initial unentangled state and for any bipartition $A,A^{\mathrm{c}}$.   Here $S_A^{\mathrm{max}} = N^\prime \log 2$, and $a \propto N^0$ is a fixed constant offset   as $N\rightarrow\infty$.      Since $\log N\approx \log N^\prime$, we will see that the choice of  $a$ or $p$ should not affect (\ref{eq:introbound}) at leading order in $N$.  Extending these definitions to quantum systems where $\mathcal{H}_i$  are higher-dimensional is straightforward.   Another generalization of the fast scrambling conjecture in more quantum information theoretic terms was given in \cite{winter}.

\section{Fast Scrambling from Local Correlation Functions}

We are now ready to answer the first of the questions posed previously: can the fast scrambling conjecture be proved for arbitrary quantum systems using purely quantum mechanical arguments? We sketch a proof here by demonstrating that scrambling is ultimately limited by the slow growth of entanglement at late times. Specifically, we shall find that $t_{\mathrm{s}} \gtrsim \log N$ because a fixed, generic $k$-local Hamiltonian cannot readily entangle two subsystems that are already nearly maximally entangled.  This slowdown of entanglement growth at late times can be related to the decay of two-point correlators, which generically decay \emph{exponentially}. While entanglement may grow rapidly at early times, this exponential time-dependence implies that the approach to maximal entanglement at long times is not arbitrarily fast.



We make this intuition precise by considering an initially unentangled set of $N$ qubits evolving under an arbitrary $k$-local Hamiltonian, and observe the approach to scrambling from the point of view of individual qubits $i$. (These arguments are easily generalized to DOF with Hilbert spaces of arbitrary dimension.) Close to the scrambling time $t_s$ a typical qubit $i$ will be highly entangled with the rest of the system, so that its reduced density matrix $\rho_i$ is nearly maximally mixed:
\begin{equation}
    \rho_i(t) = \frac{1}{2} I_i + \frac{1}{2} \sum_{\alpha=1}^3 \epsilon_i^{\alpha}(t) \sigma^{\alpha}_i
    \label{eq:SingleSiteRho}
\end{equation}
where $\epsilon_i^{\alpha}(t)$ are parameters encoding the approach of the density matrix toward the maximally mixed state $\rho_i(t) \rightarrow \frac{1}{2} I_i$.   Since the qubits are initially unentangled, $\rho_i(0)$ is pure: as a consequence, $\max_{\alpha} (\epsilon_i^\alpha(0)) \ge \frac{1}{\sqrt{3}}$.  The fast scrambling conjecture follows from two assertions:  firstly, that the functions $\epsilon_i^{\alpha}(t)$ must decay to a small value of order $\mathrm{O}(1/\sqrt{N})$ at the scrambling time $t=t_{\mathrm{s}}$; and secondly, that the decay of $\epsilon_i^\alpha(t)$ can occur no faster than exponentially quickly:  $\epsilon_i^\alpha(t) \gtrsim \mathrm{e}^{-\lambda_2 t} $ as $t\rightarrow \infty$, for a finite rate $\lambda_2 \propto N^0$.   Together, these statements imply the fast scrambling conjecture:  it is impossible to scramble at time $t \ll \lambda_2^{-1}\log N $.

The first statement is proved by considering the formal definition of scrambling (\ref{eq:SAmax}), which requires that the entanglement entropy $S_A$ of any subsystem $A$ be nearly maximal. By the property of sub-additivity, the entanglement entropy of $A$ can be no larger than the sum of the entanglement entropies of its constituent qubits:
\begin{equation}
    S_A(t) \leq \sum_{i \in A} S_i(t)
    \label{eq:Subadditivity}
\end{equation}
 where $S_i$ denotes the entanglement entropy of qubit $i$ with the other $N-1$ qubits.   Intuitively, (\ref{eq:Subadditivity}) holds because the right hand side measures entanglement between different qubits in $A$ that will not be counted in $S_A$. Using (\ref{eq:SingleSiteRho}), we can then quantify the growth of single-qubit entanglement $S_i(t)$ in terms of the decaying functions $\epsilon_i^{\alpha}(t)$.  Taylor expanding $S_i(t) = -\mathrm{tr}[\rho_i(t)\log \rho_i(t)]$, 
\begin{equation}
    S_i(t) = \log 2 - \frac{1}{4} \sum_{\alpha=1}^3 \left( \epsilon_i^{\alpha}(t) \right)^2 + \mathrm{O}\left( \epsilon^3 \right).
    \label{eq:EntropyGrowth}
\end{equation}
This expansion is sensible for $\epsilon_i^\alpha(t) \ll 1$, at a time when the initially unentangled qubits become significantly entangled.  Combining (\ref{eq:Subadditivity}) and (\ref{eq:EntropyGrowth}), we obtain:
\begin{equation}
    S_A(t) < N' \log 2 - \frac{1}{4} \sum_{i \in A} \sum_{\alpha=1}^3 \left( \epsilon_i^{\alpha}(t) \right)^2 + \mathrm{O}\left( \epsilon^3 \right).  \label{eq:SAt}
\end{equation}
In order for $S_A(t)$ to be nearly maximal as defined in (\ref{eq:SAmax}), we require that $\sum_{i,\alpha} \left( \epsilon_i^{\alpha}(t) \right)^2 < 4 a$ or  that $\epsilon_i^{\alpha}(t) \le  \sqrt{\frac{8a}{N}}$ for at least half of the qubits $i$.  When $N\rightarrow \infty$, $\mathrm{O}(\epsilon^3)$ corrections to (\ref{eq:SAt}) are negligible.

It remains to demonstrate that the functions $\epsilon_i^{\alpha}(t)$ cannot decay arbitrarily quickly, or equivalently, that entanglement cannot build up in the system arbitrarily quickly. This is the main physical content of the fast scrambling conjecture. To show this, we relate the decay of the functions $\epsilon_i^{\alpha}(t)$ to the decay of two-point correlation functions, and argue on general grounds that such correlators cannot decay arbitrarily quickly. We note that other recent work \cite{magan, stanford1802, shenker1803, lensky} has also discussed contributions to scrambling arising from the decay of two-point functions.

Using (\ref{eq:SingleSiteRho}), we may write:
\begin{equation}
    \epsilon_i^{\alpha}(t)^2 = \left( \mathrm{tr} \left[ \rho_i(t) \sigma_i^{\alpha} \right] \right)^2 = \left( \mathrm{tr} \left[ \rho \ \sigma_i^{\alpha}(t) \right] \right)^2
    \label{eq:EpsToCorrs}
\end{equation}
where $\rho$ is the initial (separable) state of the entire many-qubit system and $\mathrm{tr}$ denotes a trace over the entire system.   We expect that single-site correlation functions for the initial state $\rho$ can decay no faster than they decay for the initial state $\rho_i(0) \otimes \rho^{\infty}_{-i}$, where $\rho^{\infty}_{-i} = I_{-i} / 2^{N-1}$ is the maximally-mixed state of all qubits other than qubit $i$. That is, for a constant $C^\prime\propto N^0$,
\begin{equation}
    \left( \mathrm{tr} \left[ \rho \ \sigma_i^{\alpha}(t) \right] \right)^2 \ge C^\prime \left( \mathrm{tr} \left[ \left( \rho_i(0) \otimes \rho^{\infty}_{-i} \right) \sigma_i^{\alpha}(t) \right] \right)^2.
    \label{eq:DecayCompareToInfTemp}
\end{equation}
The intuition behind this inequality is that the state $\rho^{\infty}_{-i}$ serves as an infinite-temperature bath that relaxes qubit $i$ faster than any other initial state. A more formal perspective is found in \emph{Supporting Information}. Combining (\ref{eq:EpsToCorrs}) and assuming (\ref{eq:DecayCompareToInfTemp}) holds exactly as an inequality:
\begin{equation}
    \epsilon_i^{\alpha}(t)^2 \ge \left( \frac{1}{2^N} \sum_{\beta=1}^3 \mathrm{tr} \left[ \sigma_i^{\alpha}(t) \sigma_i^{\beta} \right] \epsilon_i^{\beta}(0) \right)^2 \ge C \mathrm{e}^{-2\lambda_2 t}  \label{eq:2lambda2}
\end{equation}
where $\lambda_2$ is the decay rate of infinite-temperature two-point correlators, and $C$ is an O(1) constant.   
Crucially, the decay rates $\lambda_2$ will be \emph{finite} for generic qubit $i$, with generic Hamiltonian.  In \emph{Supporting Information}, we find that  $\lambda_2 \propto N^0$ whenever the Hamiltonian $H$ is extensive and $k$-local (with $k$ finite), and when the density of eigenvalues of $H$ is a smooth function in the large-$N$ limit.  

Combining (\ref{eq:SAt}) and (\ref{eq:2lambda2}) with the definition of the scrambling time (\ref{eq:SAmax}), we find that  $2\lambda_2 \ge \lambda_{\mathrm{s}}$.   Entanglement grows slowly at late times whenever local two-point functions decay exponentially (which generically does occur).   Hence we derive the logarithmic lower bound (\ref{eq:introbound}) on scrambling.  The origin of the logarithmic divergence in the scrambling time is the slow saturation to a maximally entangled state at late times.   



\section{Out-of-Time-Ordered Correlators and Operator Growth}

We now turn to our final question: when might (\ref{eq:introbound}) be saturated?   The exponential decay of two-point functions is necessary but not sufficient for fast scrambling:  for example, a system made of two decoupled fast scramblers is not itself a fast scrambler (information cannot spread between decoupled systems).    To answer this question, therefore, we now turn to another probe of fast scrambling:  the growth of operators \cite{lensky, lashkari, hosur}.   In particular, we bound the growth of operators, and $t_{\mathrm{s}}$, by how the Hamiltonian connects the $N$ degrees of freedom.

Recall that in the Heisenberg picture, an operator $\mathcal{O}$ evolves in time as $\mathcal{O}(t) = \mathrm{e}^{\mathrm{i}Ht} \mathcal{O} \mathrm{e}^{-\mathrm{i}Ht}$.   Suppose that at time $t=0$,  $\mathcal{O}$ acts only on qubits in $A$:  we may store quantum information by preparing the system in the mixed state $\rho(0) = |\Psi_{\mathcal{O}}\rangle\langle \Psi_{\mathcal{O}}| \otimes 2^{-|A^{\mathrm{c}}|} I_{A^{\mathrm{c}}}$, where $|\Psi_{\mathcal{O}}\rangle$ is an eigenvector of $\mathcal{O}$.   Whereas initially $S_A[\rho(0)] = 0$,   in a scrambling quantum system $S_A[\rho(t)]$ becomes large at late times:  equivalently, almost every eigenvector of $\mathcal{O}(t)$ must be highly entangled between $A$ and $A^{\mathrm{c}}$, even though $\mathcal{O}$ acts only on $A$.  This is only possible if $\mathcal{O}(t)$ itself is a complicated operator.  We may parameterize the growth of the operator $\mathcal{O}(t)$ by writing \begin{equation}
\mathcal{O}(t) = \sum_{R\subseteq\lbrace 1,\ldots, N\rbrace}   a_R(t) \tilde{\mathcal{O}}_R(t),  \label{eq:Ovt}
\end{equation}
where $\tilde{\mathcal{O}}_R(t)$ is an operator that acts non-trivially on qubit $i$ if and only if $i\in R$, with $\lVert \tilde{\mathcal{O}}_R \rVert = 1$, where $\lVert \mathcal{O}\rVert^2 = 2^{-N} \mathrm{tr}(\mathcal{O}^\dagger\mathcal{O})$ is the operator norm. Note that the real coefficients $a_R(t)$ obey $\sum_R a_R(t)^2 = 1$.   Whereas at $t=0$, $a_R(t)\ne 0$ only when $R\subseteq A$, at the scrambling time almost all weight $a_R(t)$ is on large operators where $R$ contains $\mathrm{O}(N)$ qubits in both $A,A^{\mathrm{c}}$.   


One way to effectively describe the growth of operators without listing every $a_R(t)$ is to ask what fraction of the operator $\mathcal{O}(t)$ acts non-trivially on qubits $j$ in $A^{\mathrm{c}}$. Expanding the operators $\mathcal{O}(t),\tilde{\mathcal{O}}_R(t)$ over the complete basis of tensor products of Pauli matrices, we find that the overlap with qubit $j$ is \cite{lucas1809} 
\begin{equation}
\frac{1}{8}\sum_{\alpha=1}^3\lVert [\mathcal{O}(t),\sigma^\alpha_j] \rVert^2 = \sum_{R: j\in R} a_R(t)^2.  \label{eq:aRbound}
\end{equation}
Using (\ref{eq:aRbound}), we can easily find an operator $\sigma^\alpha_j$ for which \begin{equation}
\lVert [\mathcal{O}(t),\sigma^\alpha_j] \rVert^2 \ge \frac{8}{3} \sum_{R: j\in R} a_R(t)^2.
\label{eq:aRbound2}
\end{equation}
Since operator growth is bounded from above by (\ref{eq:aRbound2}), if we can place an upper bound on the time $t_*$ at which $\lVert [\mathcal{O}(t_*),\sigma_j^{\alpha}] \rVert$ becomes O(1), then we bound the scrambling time.  We achieve this  by generalizing \cite{lashkari, sims, hastings} the Lieb-Robinson theorem \cite{liebrobinson} to more general Hamiltonians such as (\ref{eq:introH}).   Historically, the Lieb-Robinson theorem was used to show that commutator norms in lattice models can only be large inside of an emergent ``lightcone.''    Here, we will strengthen the Lieb-Robinson theorem and show that the growth of operators is intimately tied to the structure of the interaction graph of the Hamiltonian.

For  simplicity, we will focus on 2-local Hamiltonians, leaving a more general discussion to \emph{Supporting Information}.    To any 2-local Hamiltonian
\begin{equation}
    H = \sum_{(u, v)\in E} H_{uv},    \label{eq:HvHuv}
\end{equation}
we may associate a discrete, undirected graph $G = (V,E)$, where DOF live on the vertices $v \in V$ and interact pairwise via couplings $H_{uv}$ if and only if the pair $u,v$ is connected by an edge $e = (u,v) \in E$. The connectivity of the graph $G$ is described by the \emph{adjacency matrix}
\begin{equation}
    A_{uv} = \left\lbrace \begin{array}{ll} 1 &\  (u,v) \in E \\  0 &\ (u,v)\notin E\end{array}\right.
\end{equation}
and the \emph{degree matrix} $D_{uv} = k_v \mdelta_{uv}$, where the \emph{degree} $k_v = \sum_u A_{uv}$ of a given vertex $v$ counts the number of vertices it is connected to.   Denoting \begin{equation}
\frac{c}{K} = \max_{u,v} \left( \lVert H_{uv} \rVert \right),  \label{eq:maincdef}
\end{equation}
where $K=\frac{1}{N}\sum_v k_v$ is the mean degree, and assuming $c\propto N^0$ by extensivity, our generalized Lieb-Robinson bound is  
\begin{equation}
\frac{\lVert [\mathcal{O}_u,  \mathcal{O}_v(t)] \rVert}{2\lVert \mathcal{O}_u \rVert\lVert \mathcal{O}_v \rVert} \le  \exp\left[\frac{2 c|t|}{K}(D+A)\right]_{uv} \label{eq:LR1st}
\end{equation}
where $\mathcal{O}_v$ is any operator with support only on vertex $v$. This result is derived by using the triangle inequality: $\lVert [\mathcal{O}_u, \mathcal{O}_v(t)] \rVert \le \lVert [\mathcal{O}_u, \mathcal{O}_v] \rVert + t \lVert [\mathcal{O}_u, [H,\mathcal{O}_v]] \rVert + \mathrm{O}(t^2)$.   Commutators between operators acting on disjoint sets cancel, so if $u\ne v$, $\lVert [\mathcal{O}_u, [H,\mathcal{O}_v]] \rVert = \lVert [\mathcal{O}_u, [H_{uv},\mathcal{O}_v]] \rVert $.    Furthermore, $\lVert [\mathcal{O}_u, [H_{uv},\mathcal{O}_v]] \rVert \le 4 \lVert \mathcal{O}_u \rVert \lVert H_{uv} \rVert \lVert \mathcal{O}_v \rVert$.  These bounds can be repeated at every order in $t$ and resummed to give (\ref{eq:LR1st}): see \emph{Supporting Information}. Crucially, since the result (\ref{eq:LR1st}) bounds the operator norm $\lVert \cdot \rVert$, it bounds operator growth at \emph{infinite temperature} since $\lVert \mathcal{O}\rVert^2 = \mathrm{tr}(\rho^{\infty}\mathcal{O}^\dagger \mathcal{O})$ where $\rho^{\infty} = 2^{-N} I_N$ is the infinite-temperature thermal state of all $N$ qubits.

It is instructive to average (\ref{eq:LR1st}) over all choices of $u$ and $v$: \begin{equation}
\sum_{u,v} \frac{1}{N^2} \frac{\lVert [\mathcal{O}_u,  \mathcal{O}_v(t)] \rVert}{2\lVert \mathcal{O}_u \rVert\lVert \mathcal{O}_v \rVert}  \le \frac{1}{N} \exp\left[\frac{4ck_{\mathrm{max}}}{K} |t| \right] \label{eq:expgrowth}
\end{equation}
where $k_{\mathrm{max}}$ is the maximal degree in the  graph.   Whenever $k_{\mathrm{max}}/K$ is finite, we see that operator growth is constrained to be exponentially fast. The form of the bound (\ref{eq:expgrowth}) is particularly useful because it naturally bounds the growth of \emph{out-of-time-ordered correlators} (OTOCs) such as $\mathrm{tr}(\rho [\mathcal{O}_u,  \mathcal{O}_v(t)]^2) \sim \frac{1}{N} \mathrm{e}^{\lambda_{\mathrm{L}}t}$ which measure the growth of many-body chaos with Lyapunov exponent $\lambda_L$ \cite{shenker13, douglasweak, patel, debanjan, erez, scopelliti}. Our Lieb-Robinson bound (\ref{eq:expgrowth}) establishes that the growth rate of infinite-temperature OTOCs can be no faster than 
\begin{equation}
\lambda_{\mathrm{L}} \le \frac{8ck_{\mathrm{max}}}{K},
\end{equation}  
thereby placing a bound on quantum many-body chaos at infinite temperature. Moreover, since (\ref{eq:aRbound2}) implies that scrambling may only occur once $\lVert [\mathcal{O}_u,  \mathcal{O}_v(t)]\rVert$ has grown to be O(1), we conclude from (\ref{eq:expgrowth}) that  $\lambda_{\mathrm{s}} \le \lambda_{\mathrm{L}}$.   On any graph where $k_{\mathrm{max}}/K$ is finite, we therefore have an alternative proof of the fast scrambling conjecture.

Our Lieb-Robinson bounds also constrain which graphs are capable of supporting fast scrambling. Further manipulations to (\ref{eq:LR1st}) (see \emph{Supporting Information}) lead to  \begin{equation}
\exp\left[\frac{2c|t|}{K}(D+A)\right]_{uv} < \exp\left[\frac{4\mathrm{e} ck_{\mathrm{max}}}{K} |t|  - d_{uv}\right],  \label{eq:LRdistmain}
\end{equation}
where  $\mathrm{e}\approx2.718$ and $d_{uv}$ is the minimal path length between vertices $u$ and $v$.  This is the classic Lieb-Robinson bound \cite{liebrobinson}.   Operators spread at most ballistically through the graph, and the timescale required for an arbitrary operator to spread throughout the entire system is limited by the path length between the two most distant sites $\max_{u,v}(d_{uv})$, or the \emph{graph diameter}. We immediately conclude that only graphs with diameter $\max_{u,v}(d_{uv}) \lesssim \log N$ can support fast scrambling \cite{lashkari, hastings}.
Fast scrambling is thus impossible for spin models on regular lattices in $D$ spatial dimensions \cite{prosen}, where the diameter $\propto N^{1/D}$.     On the other hand, fast scrambling is expected to be rather generic on almost every connected sparse random graph, whose diameter is $\mathrm{O}(\log N)$ for any $K>2$ \cite{bollobas}. How ``random'' must a sparse graph of fixed degree $k_v = K$ be to host a fast scrambler?   It is likely that any finite amount of non-locality is sufficient.  For example, consider the small world network \cite{watts} depicted in Figure 1d, which is constructed by starting with a one-dimensional cyclic graph where each of the $N$ vertices connects to its nearest $K/2$ neighbors on each side, and randomly re-wiring a fraction $p \ll  1/K$ of the edges to connect two randomly chosen vertices.  The typical distance between two vertices in the small world graph is \cite{newman} \begin{equation}
\frac{1}{N^2}\sum_{u,v=1}^N d_{uv} \ \approx \  \frac{\log N}{4Kp}.
\end{equation}
Using (\ref{eq:LRdistmain}), we see that  infinitesimal randomness ($p>0$) is sufficient for logarithmic scrambling.       

In order to check our assertion that fast scrambling is achievable on generic sparse graphs, we study a maximally chaotic toy model on a random graph by choosing the $H_{uv}$ in (\ref{eq:HvHuv}) to be generalized Sachdev-Ye-Kitaev (SYK) models \cite{sachdevye, kitaevunpublished, sachdev15, maldacena2016remarks}.  The SYK model is a model of $M \gg 1$ Majorana fermions interacting via random all-to-all $q$-body interactions. To construct our toy model, we select a graph $G$ consisting of $L$ vertices and place one SYK model on each vertex $v$.  The local Hilbert space on each vertex has dimension $\dim(\mathcal{H}_v) = 2^{M/2}$.  We then couple pairs of vertices $u,v$ by introducing random $q$-body interactions involving $q/2$ fermions from each of the two vertices $u,v$ whenever there is an edge $(u,v)$ in $G$.  The result is a model of $N = ML/2$ strongly-interacting degrees of freedom that is both maximally chaotic and analytically tractable. Our main result for these generalized SYK models is the computation of OTOCs; as discussed earlier, the time $t_*$ when OTOCs become large is a lower bound on $t_{\mathrm{s}}$.    We find that whenever the graph Laplacian $\Lambda_{uv} = D_{uv}-A_{uv}$ has a single vanishing eigenvalue as $N\rightarrow \infty$:
\begin{equation}
t_* = \frac{\log M + \log L}{\lambda_{\mathrm{L}}^{\mathrm{SYK}}} + \mathrm{O}(1),  \label{eq:tstarSYK}
\end{equation}
where $\lambda_{\mathrm{L}}^{\mathrm{SYK}}$ is the Lyapunov exponent of the fully connected SYK model, which has previously been computed \cite{maldacena2016remarks}.   As $\log M + \log L \approx \log N$, (\ref{eq:tstarSYK}) is the same time that it takes for OTOCs to grow large in a fully connected SYK model of $N$ fermions.    Remarkably, (\ref{eq:tstarSYK}) holds whenever the graph has the property that for an arbitrary subset $A\subset V$ with $|A|\gg 1$, the ``surface area'' of the subregion (i.e. the number of vertices in $A$ connected to vertices not in $A$) is proportional to $|A|$  (see \emph{Supporting Information}, and Figure 1).   Generic random graphs have this property, implying that all-to-all connectivity in $G$ can be redundant and is not necessary for fast scrambling.

\section{Random Quantum Circuits}
Now, let us  return to the question of OTOC growth when $k_{\mathrm{max}}/K$ diverges as $N\rightarrow \infty$.   In this regime, (\ref{eq:LR1st}) and (\ref{eq:expgrowth}) do not rule out rapid OTOC growth in a time $t_*$ independent of $N$ (or worse).    A simple example is the Hamiltonian \begin{equation}
H = \sum_{i=1}^{N-1} h_{iN}^{\alpha\beta} \sigma^\alpha_i \sigma^\beta_N. \label{eq:sigmazstar}
\end{equation}
For this model, (\ref{eq:LR1st}) implies that \begin{equation}
\frac{\lVert [\mathcal{O}_1,  \mathcal{O}_2(t)] \rVert}{2\lVert \mathcal{O}_1 \rVert\lVert \mathcal{O}_2 \rVert} \le \frac{N-1 + \mathrm{e}^{2cNt} - N \mathrm{e}^{2ct}}{N(N-1)}.   \label{eq:superNLR}
\end{equation}
In this case, the bound (\ref{eq:LR1st}) allows the possibility that OTOCs are large by the time $t_* \sim N^{-1}\log N$.

Is (\ref{eq:LR1st}) accurately capturing incredibly rapid operator growth, or is it simply a lousy upper bound on an exponentially growing commutator norm?   We can answer this question explicitly in a solvable quantum dynamical system: a random unitary circuit (RUC) \cite{hayden07, fawzi, nahum16, nahum,  tibor}.   In the RUC, the time evolution operator $\mathrm{e}^{-\mathrm{i}Ht}$ is replaced by a product of $\lceil Nt \rceil$ unitary operations $U$, each of which act on a finite number $k$ of the DOF (to mimic $k$-locality). The simplifying assumption is that we uniformly average over all allowed local unitary operations, which reduces the resulting dynamics to a classical stochastic process \cite{nahum16, nahum, tibor}.    

For example, consider a 2-local RUC in the limit where the local Hilbert space dimension $d \rightarrow \infty$. (A full discussion of RUC dynamics at finite $d$ is presented in \emph{Supporting Information}.) A 2-local unitary $U_{uw}$ acting on vertices $u$ and $w$ almost surely acts on tensor product operators as follows, where we use $\mathcal{O}$ to represent any non-identity operator:
\begin{subequations}\label{eq:RUCdynamics}\begin{align}
U_{uw}^\dagger (I_u \otimes I_w \otimes \mathcal{O}_{\mathrm{rest}}) U_{uw} &= I_u \otimes I_w \otimes \mathcal{O}_{\mathrm{rest}}, \\
U_{uw}^\dagger (I_u \otimes \mathcal{O}_w \otimes \mathcal{O}_{\mathrm{rest}}) U_{uw} &= \mathcal{O}_u \otimes \mathcal{O}_w \otimes \mathcal{O}_{\mathrm{rest}}, \\
U_{uw}^\dagger (\mathcal{O}_u \otimes I_w \otimes \mathcal{O}_{\mathrm{rest}}) U_{uw} &= \mathcal{O}_u \otimes \mathcal{O}_w \otimes \mathcal{O}_{\mathrm{rest}}, \\
U_{uw}^\dagger (\mathcal{O}_u \otimes \mathcal{O}_w \otimes \mathcal{O}_{\mathrm{rest}}) U_{uw} &= \mathcal{O}_u \otimes \mathcal{O}_w \otimes \mathcal{O}_{\mathrm{rest}}.
\end{align}\end{subequations}
Here $\mathcal{O}_{\mathrm{rest}}$ denotes operators acting on the remaining $N-2$ vertices.   Suppose that the allowed $u$ and $w$ correspond to a uniformly chosen edge from a given graph $G$.   From (\ref{eq:RUCdynamics}), if an arbitrary operator $\tilde{\mathcal{O}}$ acts either on $u$ or $w$,  then $U^\dagger_{uw} \tilde{\mathcal{O}} U_{uw}$ acts on both $u$ and $w$. A series of random unitary operations $U_{uw}$ applied in this manner will spread information from an initial vertex $v$ across the entire system, similar to the way in which an operator $\mathcal{O}_v(t)$ initially localized on vertex $v$ grows under time evolution as in (\ref{eq:Ovt}).  This is reminiscent of a spreading `infection,' where we say a vertex $u$ is `infected' if and only if a non-identity operator $\mathcal{O}_u$ is present on $u$. From (\ref{eq:RUCdynamics}), infections spread whenever we choose an edge that connects infected and uninfected vertices.   The resulting infection dynamics is equivalent to the well-known SI model \cite{vespignaniRMP}.     Therefore, we can use the theory of infection spreading on general graphs to understand the spread of operators in the RUC.

Of particular interest is the fact that infections spread faster than exponentially on heterogeneous networks in the SI model \cite{vespignani}.  A heterogeneous network is one where the probability $p(k)$ that vertex $v$ has degree $k_v=k$ has a heavy tail, with the variance of $p(k)$ diverging as $N\rightarrow \infty$.   An example is the interaction graph of (\ref{eq:sigmazstar}).   An RUC on this graph will be `super-chaotic,' with  $\lambda_{\mathrm{L}}=\infty$.    
 To model the commutator in (\ref{eq:superNLR}), we start with vertex 1 infected at $t=0$.   At time $t_0$ (whose probability distribution is $\mathbb{P}(t_0 > s) = \mathrm{e}^{-s}$) the infection will jump to the central vertex $N$.   Starting from the time $t_0$ when $N$ is infected, the probability that vertices $2,\ldots,N-1$ are infected is $1-\mathrm{e}^{t_0-t}$ ($t>t_0$).   At time $t=t_0 + \log 2$, vertex 2 is infected with probability $\frac{1}{2}$; hence $\lVert [\mathcal{O}_1(t), \mathcal{O}_2]\rVert$ is large.   Thus, this RUC exhibits an operator growth time $t_* \propto N^0 \ll \log N$.   While our bound (\ref{eq:superNLR}) is very weak, the conclusion that operator growth is faster than on a regular graph, where all vertices have the same degree, remains.   

In agreement with our earlier discussions, however, the scrambling time in the RUC is ultimately limited by the slow growth of entanglement entropy at late times.   To demonstrate this, suppose we start in a tensor product state $|\Psi(0)\rangle = \otimes_v |\Psi_v\rangle$.    Choose an arbitrary subset $A$ of half of the vertices, with $S_A(t=0)=0$.   To generate nearly maximal entanglement $S_A(t) \ge \frac{N}{2}\log d -1 $,  at least one unitary must hit every vertex $v\in A$;  otherwise, the reduced density matrix $\rho_A = |\Psi_v \rangle \langle \Psi_v | \otimes \cdots$ and we obtain $S_A \le (\frac{N}{2}-1)\log d$.   In the RUC, the probability that we have not chosen a vertex $v$ by time $t$ is $\exp[-\frac{k_v}{K}t]$, where $k_v$ is the degree of $v$ and $K$ is the mean degree.     At least half of the vertices $v$ have $k_v \le 2K$.   Therefore, the time $t_A$ it takes to choose every vertex in $A$ obeys $t_A \ge  \frac{1}{2}\log N$ with high probability:  at shorter times at least one vertex has not yet been chosen.  Since  $t_{\mathrm{s}} \ge t_A$, the fast scrambling conjecture (\ref{eq:introbound}) holds, despite the system exhibiting an infinite Lyapunov exponent.

\section{Discussion}

We have shown that the scrambling time in few-body quantum dynamics is always bounded by the decay rates of local operators.  By contrast, there is no universal relationship between fast scrambling and  quantum chaos. In particular, we emphasize that the rapid growth of OTOCs is not always a reliable indicator of  fast scrambling. For instance, the SYK model is highly chaotic, but generates entanglement extremely slowly at late times \cite{gu1708}.   On the other hand, we found a RUC model which is `super-chaotic' ($\lambda_{\mathrm{L}} = \infty$) at early times, yet still obeys the fast scrambling conjecture. Furthermore, while black holes are believed to be the fastest scramblers in nature \cite{susskind08}, they are `less' chaotic than this RUC: known Hamiltonians related to black holes have a finite Lyapunov exponent at all temperatures.    One reason for this is that microscopic quantum models of black holes \cite{bfss} have homogeneous, $k$-local, all-to-all connectivity.  Heterogeneous connectivity is required to obtain $\lambda_{\mathrm{L}}=\infty$, as shown by our generalized Lieb-Robinson bounds.

That scrambling is instead intimately connected to local decay rates is reminiscent of the historical origins of the fast scrambling conjecture in  black hole physics. The ringdown of quasinormal modes in a time $t_{\mathrm{s}} \propto \log N$ is in large part the motivation for the  conjecture \cite{hayden07, susskind08}.    Furthermore, the scrambling time  of black holes appears as fast as possible without measurable violations of quantum no-cloning theorems \cite{hayden07}.  Rather than cloning information, black holes rapidly generate entanglement between degrees of freedom. Our arguments show that the rate of this entanglement growth is limited by the finiteness of decay rates in general quantum systems. In the context of the AdS/CFT correspondence \cite{lucasbook}, these decay rates are those of black hole quasinormal modes. This makes the original justification of fast scrambling precise.

\begin{figure}[t]
\centering
\includegraphics[width=6in]{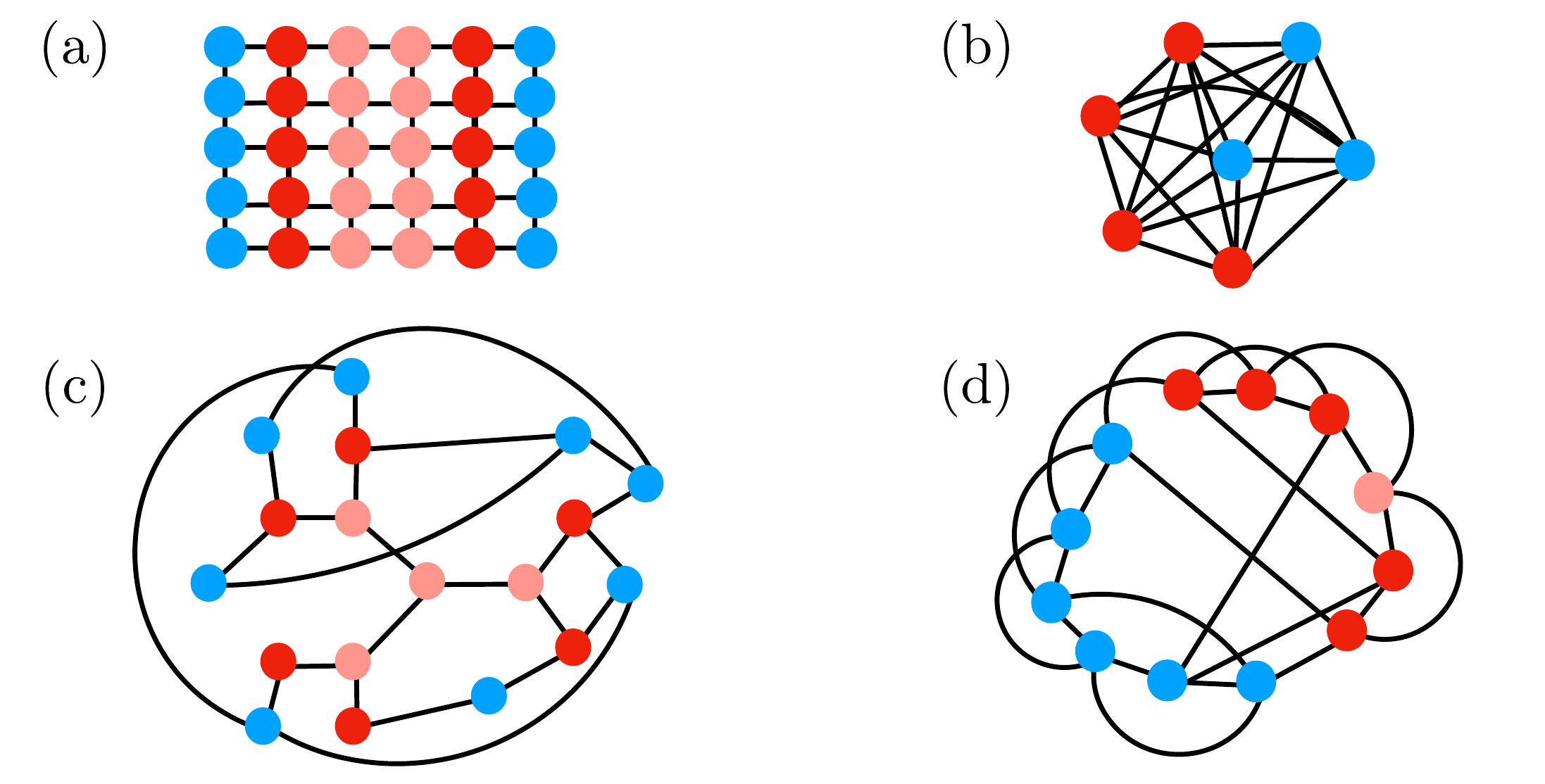}
\caption{A selection of different graphs of relevance in this paper: (a) a regular lattice; (b) a complete graph; (c) a sparse random graph which is locally treelike; (d) a small world graph.   (a) cannot host fast scramblers due to the large graph diameter, but (b)-(d) can host fast scramblers.  A subset $A$ of vertices in each graph above is denoted in red; vertices  in $A^{\mathrm{c}}$ are denoted in blue.  Light red vertices denote vertices away from the `area' of $A$, and dark red vertices are at the boundary of $A$ and $A^{\mathrm{c}}$.  In the thermodynamic limit, on graphs (b)-(d), a finite fraction of all edges starting in $A$ end in $A^{\mathrm{c}}$:  thus area is proportional to the number of DOF inside the region.}
\label{fig:graphsfig}
\end{figure}

There is another remarkable analogy between black holes and quantum dynamics on random graphs.  The area of a black hole is proportional to the thermal entropy \cite{hawking}, which is in turn proportional to the total number of degrees of freedom $N$.  Generic random graphs have a similar property, where a significant fraction of vertices in any connected region $A \subset V$ are located at the \emph{boundary} of $A$ -- they connect to the `outside.' This is in contrast to, say, a regular lattice, in which most of the vertices of any particular connected region $A$ reside deep within the volume of $A$ and do not connect to vertices outside $A$.  See \emph{Supporting Information} for a precise statement, and Figure \ref{fig:graphsfig} for a pictorial demonstration of this result; see also \cite{magan2}. As a specific example, on any graph with this property ``$\text{area}\propto N$,''  our generalized SYK model is just as chaotic as the fully connected model, where the fully connected model is itself a crude model of a black hole \cite{kitaevunpublished, maldacena2016remarks}.

\section{Outlook}

The most important extension of our results is to finite  temperature quantum dynamics.  We conjecture that at all temperatures and for all initially unentangled states, the scrambling time is limited by the decay of thermal two point functions.  A proof likely requires more sophisticated methods than what we have developed here: see \cite{lensky} for some preliminary directions.  More interestingly, our explicit construction of random circuits with $\lambda_{\mathrm{L}}=\infty$ is surprising given that, under mild assumptions, Lyapunov exponents generally are bounded from above by $\lambda_{\mathrm{L}} \le 2\mpi k_{\mathrm{B}}T/\hbar$ at finite temperature $T$ \cite{stanfordbound}, where $\hbar = h/2\mpi$ and $k_{\mathrm{B}}$ and $h$ are Boltzmann's and Planck's constants,  respectively.  Our work does not contradict \cite{stanfordbound}, however: the RUC is not generated by a fixed Hamiltonian, and should be considered as a model of infinite temperature dynamics where the bound of \cite{stanfordbound} is trivial.   However, a necessary condition to obtain $\lambda_{\mathrm{L}} \le 2\mpi k_{\mathrm{B}}T/\hbar$ is that OTOCs stay sufficiently small ($\mathrm{O}(\frac{1}{N})$) at early times.   This condition  fails on heterogeneous graphs.

Our discussion of scrambling on sparse graphs is also relevant for experimental efforts to realize highly chaotic quantum systems.   Simulations of Hamiltonians on sparse graphs can leverage, for instance, the non-local interactions naturally afforded by cavity quantum electrodynamics (QED) setups. Cavity QED systems have long provided the ability to engineer strong non-local interactions between atoms, where virtual photons in the cavity mode mediate correlated spin-flips between positionally fixed atoms whose ground states serve as pseudo-spin-$\frac{1}{2}$ degrees of freedom \cite{Miller2005, Walther2006, Ritsch2013}. While single-mode cavities naturally generate uniform all-to-all couplings between spins \cite{Sorensen2002, Strack2011, Leroux2010, Hosten2016}, additional experimental tools may be used to engineer non-local interactions on more general graphs.

Spin models on sparse graphs, for instance, can be obtained in a single-mode cavity by suppressing interactions with a magnetic field gradient and selectively enhancing interactions between pairs of atoms at particular distances by driving the ensemble with multi-frequency light \cite{Hung2016}. Similarly, by coupling an ensemble of atoms to a multimode cavity and controlling their positions with, e.g., optical tweezers \cite{Endres2016}, one may obtain spin models whose interactions correspond to a complete graph, but whose strength and sign are disordered, similar to random couplings in the SYK model \cite{Gopalakrishnan2011, Vaidya2018}. While these tools would enable the implementation of scrambling dynamics, several groups have also demonstrated techniques to measure OTOCs via interferometric protocols \cite{Swingle2016, Garttner2017, Meier2017, Li2017}, thereby allowing one to experimentally observe and characterize the growth of operators in these systems. Owing to the presence of photon losses in cavity setups, one is always restricted by the effects of dissipation which limit the allowed time evolution of the system before decoherence sets in. With state-of-the-art cavity systems deep in the strong coupling regime, however, we expect that it is possible to realistically observe scrambling in systems of as many as 100 spins in the near future.

\addcontentsline{toc}{section}{Acknowledgements}
\section*{Acknowledgements}
We thank Geoffrey Bentsen, Patrick Hayden, Yuri Lensky, Xiao-Liang Qi, Steve Shenker and Brian Swingle for helpful discussions. We also thank Tori Borish and Ana Maria Rey for providing feedback on the manuscript.
This work supported by the Gordon and Betty Moore Foundation's EPiQS Initiative through Grants GBMF4306 (YG) and GBMF4302 (AL), and by the Office of High Energy Physics of the Department of Energy (GB).

\begin{appendix}

\section{Bounding the Growth of Entanglement}
\label{sec:subsystem}
The supporting information contains the technical details of the results announced in the first section.  Henceforth, we will assume that the Hamiltonian is $k$-local on a hypergraph with $L$ vertices, and that the local Hilbert space dimension is $2^M$.   Here we are only interested in models where $M$ is strictly finite: e.g. $M=1$;\footnote{In  SYK-like models, the local Hilbert spaces below should be interpreted as few fermion clusters on a given graph vertex, in contrast to our interpretation in (\ref{eq:tstarSYK}), and in Appendix \ref{sec:SYK}.} thus $\log N \approx \log L$.

In this section, we sketch a proof that $\lambda_2$ is finite under rather general circumstances:  (\emph{i}) the $k$-local Hamiltonian (\ref{eq:introH}) is extensive, and (\emph{ii}) the spectrum of the many-body Hamiltonian approaches a smooth distribution in the thermodynamic limit.   The proof proceeds in two parts.    Let $\rho_v(t)$ be the  reduced density matrix for vertex $v$ at time $t$, and define 
\begin{equation}
S^{(1)}_v(t) = -\mathrm{tr}\left[\rho_v(t)\log \rho_v(t)\right], \;\;\;\; S^{(2)}_v(t) = -\log \mathrm{tr}\left[\rho_v(t)^2\right].
\end{equation} 
First, we show that if $M\log 2 - S^{(2)}_v = \delta$, and  $2^M\delta \ll 1$, 
\begin{equation}
\left| \left(M\log 2 - S^{(1)}_v\right) - \frac{\delta}{2}\right| \le \frac{2^{(M-1)/2}}{3} \delta^{3/2} + \mathrm{O}\left(\delta^2\right).  \label{eq:S1S2}
\end{equation}
Secondly, we show that
\begin{equation}
\frac{ \left| \langle  \mathcal{O}_v(t) \mathcal{O}^\prime_v(0)\rangle \right| }{\lVert \mathcal{O}_v \rVert\lVert \mathcal{O}_v^\prime \rVert }\ge C \mathrm{e}^{-\lambda^*_v t},  \label{eq:lambdastar}
\end{equation}
where $\mathcal{O}_v$ and $\mathcal{O}^\prime_v$ are local operators on vertex $v$, and $\lambda^*_v$ is a $v$-dependent decay rate which is independent of $N$ for almost all vertices.  

 With these two results in hand, we now consider the von Neumann entanglement entropy of a  region $A$  consisting of $L^\prime=pL$ of the vertices in the graph, with $0<p\le \frac{1}{2}$.   Since \begin{equation}
 ML^\prime \log 2 - S_A \ge \sum_{v\in A} \left[ M\log 2 -  S_v^{(1)}\right] \approx \frac{1}{2} \sum_{v\in A} \left[ M\log 2 -  S_v^{(2)}\right],
 \end{equation}
 and \begin{equation}
 M\log 2 -  S_v^{(2)} \gtrsim \mathrm{e}^{-2\lambda_{2v}t},
 \end{equation}
if half of the vertices within $A$ have decay rates $\lambda_{2v} \le \lambda_2$, then such vertices must (on average) have $\delta \le \frac{4a}{L^\prime }$ in order for $ML^\prime \log 2 - \sum S_v  \le  a$.   In the thermodynamic limit ($L\rightarrow \infty$ with $M$ finite), we will show that there exists a positive constant $C\propto N^0$ for which $\delta \gtrsim C \mathrm{e}^{-\lambda_2 t}$, which proves (\ref{eq:2lambda2}).
\subsection{Relating Renyi to von Neumann Entropy}
First  we show (\ref{eq:S1S2}).   If $0\le p_i \le 1$ denote the eigenvalues of the reduced density matrix $\rho_v$, then \begin{equation}
S^{(1)}_v  = -\sum_i p_i \log p_i, \;\;\;\; S^{(2)}_v  = -\log \left(\sum_i p_i^2\right).
\end{equation}
Also observe that $\sum p_i = 1$.  Let \begin{equation}
p_i = \frac{1+\epsilon_i}{n},
\end{equation}
and let $M\log 2 - S^{(2)}_v = \delta$.  If $2\delta < 1$, then by definition 
\begin{equation}
\frac{1}{2^{M}} \sum_i \epsilon_i^2 = \mathrm{e}^\delta-1 < \delta+\delta^2 < 2\delta.  \label{eq:sumeps2}
\end{equation}
Thus, for all $i$, \begin{equation}
|\epsilon_i| \le \sqrt{2^{M+1}\delta}. \label{eq:epsibound}
\end{equation}
Now, when $2^{M+1}\delta \ll 1$, \begin{align}
S^{(1)}_v &= M\log 2 - \frac{1}{2^{M+1}}\sum_i \epsilon_i^2  + \frac{1}{6\cdot 2^M} \sum_i \epsilon_i^3 -  \frac{1}{12\cdot 2^M} \sum_i \epsilon_i^4 + \cdots,
\end{align}
and so  \begin{equation}
\left| \left(M\log 2 - S^{(1)}_v \right) - \frac{1}{2} \frac{1}{2^{M}} \sum_i \epsilon_i^2  \right| \le \left(\frac{\max(|\epsilon_i|)}{6}  + \frac{\max(|\epsilon_i|)^2}{12}+ \cdots\right)\frac{1}{2^{M}} \sum_i \epsilon_i^2  
\end{equation}
Taylor expanding the equality in (\ref{eq:sumeps2}), and using (\ref{eq:epsibound}), we obtain (\ref{eq:S1S2}).

\subsection{Renyi Entropy and the Memory Function}
Next, we derive an exact, albeit non-local in time, equation of motion for $S^{(2)}_v(t)$.   To do so, let us consider the vector space of all Hermitian operators which act on the Hilbert space of vertex $v$.   The density matrix $|\rho_v)$ is in this vector space.   A basis set  for this vector space is $\lbrace |\mathbb{I}), |T^a_v)\rbrace$, where $T^a_v$ denote the Hermitian generators of $\mathrm{SU}(2^M)$, acting on vertex $v$ alone.   An (orthogonal) inner product for this basis is  \begin{equation}
(\mathcal{O}_1|\mathcal{O}_2) \equiv \mathrm{tr}\left(\mathcal{O}_1\mathcal{O}_2\right).
\end{equation}

Our goal is now to find a linear equation for $|\rho_v(t))$.   This can be done, as the Schr\"odinger equation is linear.   However, as we are integrating out degrees of freedom, the resulting equation will be nonlocal.   This is the essence of the memory function formalism  \cite{mori, zwanzig, forster}.   Unlike in the usual of the memory function equations (in real time), here we will also need to keep track of additional terms related to the initial conditions of the entire many-body state.   We  will ultimately argue that  these additional terms cannot generally modify (\ref{eq:lambdastar}).

The key point is as follows.    First, let us temporarily expand the vector space to include  the set of all Hermitian matrices.   Our goal is to project the dynamics back on to the set of reduced density matrices on vertex $v$, which is an exponentially small subset of all possible basis vectors.    Let the initial many-body state be $|0)$.   Since $\mathrm{tr} \rho = 1$ for all times, we know that $|0)$ and $|\mathbb{I})$ have some overlap, so we may write the initial state as \begin{equation}
|0) = |0_v) + |\tilde 0).
\end{equation}
The first term is the initial condition for $\rho_v$, tensored with the identity on all other sites;  the second term makes up the remainder of terms.   Let $\mathcal{L}$ be the Liouvillian: \begin{equation}
\mathcal{L}|\mathcal{O}) =  -|[H,\mathcal{O}]),
\end{equation}
 with $H$ the many-body Hamiltonian.   The time evolution of the density matrix is given by $\mathrm{e}^{-\mathrm{i}\mathcal{L}t}|0)$.   Denoting $\mathfrak{p}$ as the projection operator onto $\lbrace |\mathbb{I}), |T^a_v)\rbrace$, and $\mathfrak{q}=1-\mathfrak{p}$, we thus wish compute \begin{equation}
 |\rho_v(t)) = \mathfrak{p}\mathrm{e}^{-\mathrm{i}\mathcal{L}t} |0).
 \end{equation}
 More specifically, since  \begin{equation}
 \mathrm{e}^{-S^{(2)}_v(t)} = ( \rho_v(t)|\rho_v(t)),
 \end{equation}
 we wish to compute \begin{equation}
 \frac{\mathrm{d}S^{(2)}_v}{\mathrm{d}t} = - \frac{2}{( \rho_v(t)|\rho_v(t))} ( \rho_v(t) |\frac{\mathrm{d}|\rho_v(t))}{\mathrm{d}t}.
 \end{equation}
Using the identity \begin{equation}
 \mathfrak{p} \mathrm{e}^{-\mathrm{i}\mathcal{L}t}  =  \mathfrak{p} \mathrm{e}^{-\mathrm{i}\mathcal{L}\mathfrak{q}t} - \mathrm{i}\mathfrak{p} \int\limits_0^t \mathrm{d}s \mathrm{e}^{-\mathrm{i}\mathcal{L}\mathfrak{q}(t-s)}\mathcal{L} \mathfrak{p} \mathrm{e}^{-\mathrm{i}\mathcal{L}s},
 \end{equation}
 we obtain \begin{equation}
 \frac{\mathrm{d}|\rho_v(t))}{\mathrm{d}t} =   - \mathrm{i}\mathfrak{p}\mathcal{L}\mathfrak{q}\mathrm{e}^{-\mathrm{i}\mathfrak{q}\mathcal{L}\mathfrak{q} t} |\tilde 0) - \mathrm{i} \mathfrak{p} \mathcal{L}\mathfrak{p} |\rho_v(t))  - \int\limits_0^t \mathrm{d}s \; \mathfrak{p} \mathcal{L}\mathfrak{q} \mathrm{e}^{-\mathrm{i}\mathfrak{q}\mathcal{L}\mathfrak{q}s} \mathfrak{q}\mathcal{L} \mathfrak{p} |\rho_v(t-s)). \label{eq:generaleqsec2}
 \end{equation}
 This is a linear equation (albeit nonlocal in time) with a source, given by the first term on the right hand side.  We may write its solution as the sum of a particular solution and a homogeneous solution: \begin{equation}
 |\rho_v(t)) = |\bar\rho^0_v(t)) + |\tilde \rho_v(t)) 
 \end{equation} 
 The first term above is a homogeneous solution of (\ref{eq:generaleqsec2}), obeying $\mathfrak{p}|\bar \rho^0_v)= |\bar \rho^0_v)$; the second term is a particular solution of (\ref{eq:generaleqsec2}) obeying $|\tilde\rho_v(0)) = 0$.   For generic initial conditions, we expect the decay of $|\rho_v(t))$ to $\sim |\mathbb{I})$ can be no faster than the decay of the homogeneous term.    So we expect that it suffices to bound the decay of $|\bar\rho^0_v(t))$.   This is the first point at which we will sacrifice some rigor and simply sketch out a proof of (\ref{eq:lambdastar}).   The fact that a generic linear equation decays no faster than the decay of the homogeneous solution is the formal version of the inequality we provided in (\ref{eq:2lambda2}).
 
 It is most natural to describe the dynamics of $|\bar\rho^0_v(t))$ via a Laplace transform.    It is convenient to subtract out the identity component: \begin{equation}
 |\bar\rho_v(t)) \equiv |\bar\rho^0_v(t)) - \frac{1}{2^M} |\mathbb{I}).
 \end{equation}
 Since $|\mathbb{I})$ is a null vector of $\mathcal{L}$, it decouples  from the dynamics and we will  omit this basis vector in what follows.  The assumption that entropy saturates to maximal then implies that $\bar\rho_v(t)) \rightarrow 0$ as $t\rightarrow \infty$.
 Letting \begin{equation}
 |\bar\rho_v(z)) \equiv \int\limits_0^\infty \mathrm{d}t\; \mathrm{e}^{-zt} |\rho_v(t)),
 \end{equation}
 we find that \begin{equation}
 z|\bar\rho_v(z)) - |\rho_v(0)) = -\left(\mathrm{i}\mathfrak{p}\mathcal{L}\mathfrak{p} + \mathcal{K}(z) \right) |\bar\rho_v(z)),
 \end{equation}
 where the memory function \begin{equation}
 \mathcal{K}(z)  \equiv \mathfrak{p} \mathcal{L}\mathfrak{q} \left(z+\mathrm{i}\mathfrak{q}\mathcal{L}\mathfrak{q}\right)^{-1} \mathfrak{q}\mathcal{L} \mathfrak{p}.
 \end{equation}
 The inverse Laplace transform then gives us the time-dependent reduced density matrix: \begin{equation}
 |\bar\rho_v(t))  = \int\limits_{z_0+\mathrm{i}\mathbb{R}}  \frac{\mathrm{d}z}{2\mpi\mathrm{i}}  \;   \mathrm{e}^{zt}  \left(z + \mathcal{K}(z) + \mathrm{i}\mathfrak{p}\mathcal{L}\mathfrak{p}\right)^{-1} |\rho_v(0)) 
 \end{equation}
 The real number $z_0$ is chosen somewhere where the contour can be closed: i.e. for $z_0\le  \mathrm{Re}(z_*)$, where $z_*$ is the complex number with smallest real part such that $z+\mathcal{K}(z)+\mathrm{i}\mathfrak{p}\mathcal{L}\mathfrak{p}$  is a singular matrix.   For  the moment, let us assume that $z_*$ is finite.  We will justify this in the next subsection, together with $\mathrm{Re}(z_*)<0$.  Standard theorems then  give that (for generic initial conditions) as $t\rightarrow \infty$ \begin{align}
 |\bar\rho_v(t)) &\approx\mathfrak{p}_* \int\limits_{z_0+\mathrm{i}\mathbb{R}}  \frac{\mathrm{d}z}{2\mpi\mathrm{i}}  \;   \mathrm{e}^{zt}  \left(z + \mathcal{K}(z) + \mathrm{i}\mathfrak{p}\mathcal{L}\mathfrak{p}\right)^{-1}\mathfrak{p}_* |\rho_v(0)) = \int\limits_{z_0+\mathrm{i}\mathbb{R}}  \frac{\mathrm{d}z}{2\mpi\mathrm{i}}  \; \frac{\mathrm{e}^{zt}}{z-z_*} C_* |\rho_v(0)) = \mathrm{e}^{z_*t} C_* |\rho_v(0))
 \end{align}
 where $\mathfrak{p}_*$ projects onto all vectors with this most singular `eigen'vector, and $C_* $ is a suitable matrix.
 Subleading corrections to this equation decay with a faster exponential  rate.   We then  conclude that (\ref{eq:lambdastar}) holds, with $\lambda^*_v = -2\mathrm{Re}(z_*)$ and $C = (\rho_v(0)| C_*^2 |\rho_v(0))$.

\subsection{Finiteness of the Memory Function}
It remains  to justify why $-\infty < \mathrm{Re}(z_*)<0$.   First, let us suppose that $|E_\alpha)$ denote the eigenfunctions of $\mathfrak{q}\mathcal{L}\mathfrak{q}$ in the image of $\mathfrak{q}$.  We write
\begin{equation}
\mathfrak{p}\mathcal{L}\mathfrak{q} = \sum_\alpha   c_{\alpha E_\alpha}  |\psi_\alpha)  (E_\alpha|
\end{equation}
Let us further assume that the density of states of the many-body Hamiltonian is continuous, and is given by $\rho(E)$.     We then find that the density of states of both $\mathcal{L}$ and $\mathfrak{q}\mathcal{L}\mathfrak{q}$ is given by \begin{equation}
\rho_{\mathcal{L}}(E) = \int \mathrm{d}E^\prime \; \rho(E^\prime) \rho(E^\prime+E).
\end{equation}
Because $\mathfrak{q}\mathcal{L}\mathfrak{q}$ and $\mathcal{L}$  are identical matrices up to an exponentially small number of entries, corrections to their relative continuous spectra are $\propto \mathrm{e}^{-L}$, and such effects are comparable, in the thermodynamic limit, to other ``finite size" effects that we neglect.  We further assume that \begin{equation}
\rho_{\mathcal{L}}(E) \mathcal{F}(E)  \equiv \sum_{\alpha} \mdelta(E_\alpha - E)  |c_{\alpha E_\alpha}|^2  |\psi_\alpha)  (\psi_\alpha|   \label{eq:FEeq}
\end{equation}
is a smooth-valued positive definite matrix as a function of $E$, in the thermodynamic limit.\footnote{It is, by  definition, positive semidefinite;  we further assume that there are no null vectors.}    If these two assumptions hold, we obtain \begin{equation}
\mathcal{K}(z) = \int \mathrm{d}E\;   \frac{\rho_{\mathcal{L}}(E) \mathcal{F}(E)}{z-\mathrm{i}E}
\end{equation}

We first show that $-\infty < \mathrm{Re}(z_*)$.   We do this by bounding $\mathcal{F}$, since  \begin{equation}
(\alpha| \mathcal{K}(z) |\alpha) < \mpi \max_E \left((\alpha|\rho_{\mathcal{L}}(E) \mathcal{F}(E)|\alpha).\right).
\end{equation}
First, define \begin{equation}
\mathcal{L}_v = \mathcal{L}\mathfrak{p}.
\end{equation}
For a $k$-local Hamiltonian, using the $\mathrm{L}_\infty$ operator norm (for simplicity): \begin{equation}
\frac{1}{L}\sum_{v\in V} \lVert \mathcal{L}_v  \rVert \le \frac{2k\lVert H \rVert}{L} \propto L^0.   \label{eq:boundL}
\end{equation}
The last step uses extensivity.   Using Markov's inequality, this means that a finite fraction of vertices have $\lVert \mathcal{L}_v \rVert < M_0$, for an $L$-independent constant $M_0$ (as $M_0 \rightarrow \infty$, the fraction goes to 1).  From (\ref{eq:FEeq}), we find \begin{equation}
(\alpha| \mathcal{F}(E)  |\alpha) <  (\alpha| \mathfrak{p} \mathcal{L}_v^2 \mathfrak{p} | \alpha) <  M_0^2  (\alpha|\alpha).
\end{equation}
Hence\begin{equation}
\mathrm{Re}(z_*) > -M_0^2 \max_E \rho_{\mathcal{L}}(E) > -\infty.
\end{equation}


To argue that $\mathrm{Re}(z_*)<0$, observe that  \begin{equation}
\mathrm{Re}(\mathcal{K}(0)) = \mpi \mathcal{F}(0) \int \mathrm{d}E \; \rho(E)^2.
\end{equation}
 $\mathrm{Re}(z_*)<0$ whenever $\mathcal{F}(0)$ is  positive definite.   It is positive semidefinite by construction.  We do expect that all eigenvalues are strictly positive in a typical chaotic model.  One subtlety arises from the possiblity of Hamiltonians which have  continuous many-body spectrum but are integrable.   These include either the quantum version of ``classical"  Ising Hamiltonians with integer coefficients, or many-body localized systems \cite{rahul}.    For these systems, we do expect that $\mathcal{F}$ has null vectors or vectors that are so close to null that the above proof fails.  For example, in the many-body localized theories, one can generalize the above procedure to finite subsets of vertices.  Localization of eigenstates will apply that some eigenvalues of $\mathcal{F}$ are extremely small, when the subset of vertices is finite in the thermodynamic limit, but large compared to the localization length.
 
Subject to the caveats described above, this completes the argument for the finiteness of $\lambda_v^*$ for most vertices.   Our assumptions seem quite plausible in chaotic many-body systems.  As stated in the main text, it would be interesting if the ``proof" sketched here can be made rigorous.

\subsection{Possible Generalizations}
We now briefly comment on a few possible generalizations of our results.    Firstly, we  expect that the bounding of entanglement generation at late times by the decay of correlation functions also holds at finite temperature, though a generalization of our ``proof" appears non-trivial.  Nevertheless, we present some circumstantial evidence that hints towards a finite temperature generalization of (\ref{eq:2lambda2}).   Combining results from \cite{solodukhin} and \cite{shenker13}, we find that scrambling at late times is set by the decay rate of two point functions at finite temperature in two dimensional holographic conformal field theories.   Another example where the (partial) saturation of entanglement is related to the decay of two-point functions is the Sachdev-Ye-Kitaev model \cite{gu1708}.

 The presence of conservation laws, including energy conservation, may further slow down the growth of entanglement by preventing the rapid decay of mutual information between vertices within the subregion $A$ \cite{tibor2, khemani}.   Such effects further increase $t_{\mathrm{s}}$ beyond $\log L / (2\lambda_2)$, consistent with the fast scrambling conjecture.   

The growth of entanglement is extremely rapid at early times, and slows down at  late times.\footnote{Although saturation of entanglement entropy after a quench appears to be abrupt in translation invariant  theories with a holographic dual \cite{suh1, suh2},  such models do not saturate (\ref{eq:introbound}): the spatial spread of information is ``slow". }   This is consistent with bounds on the rate of entanglement generation, which show that the generation of entanglement is proportional to the perimeter of $A$ \cite{childs, bravyi, marien}.   So we can formally  bound the growth of entanglement at early times.  For simplicity, we will focus on a 2-local Hamiltonian with operator norms (\ref{eq:maincdef}), though we will now let $\alpha,\beta$ run from $1$ to $2^{2M}-1$: \cite{bravyi, marien} \begin{equation}
\frac{\mathrm{d}S_A}{\mathrm{d}t} < 18M\log 2 \times \frac{c}{K} \times |E_{A,A^{\mathrm{c}}}| < 18 cN\log 2.
\end{equation}
In the above equation, $|E_{A,A^{\mathrm{c}}}|$ counts the number  of  edges between $A$ and $A^{\mathrm{c}}$.  On a generic, locally tree-like graph, nearly every vertex is at the edge of $A$ \cite{bollobas}, so we do expect $\mathrm{d}S_A/\mathrm{d}t\sim N$ at early times.    The fact that $\mathrm{d}S_A/\mathrm{d}t \propto N$ at early times was also noted in \cite{magan}; however, in \cite{magan}, $a\propto N$ in (\ref{eq:SAmax}), so that  $t_{\mathrm{s}} \propto N^0$ is possible.  We view this $N$-independent time scale as a time scale for local thermalization, but not for scrambling.

\section{Lieb-Robinson Bounds}
\label{app:LR}

\subsection{Derivation}
\begin{figure}
\centering
\includegraphics[width=5in]{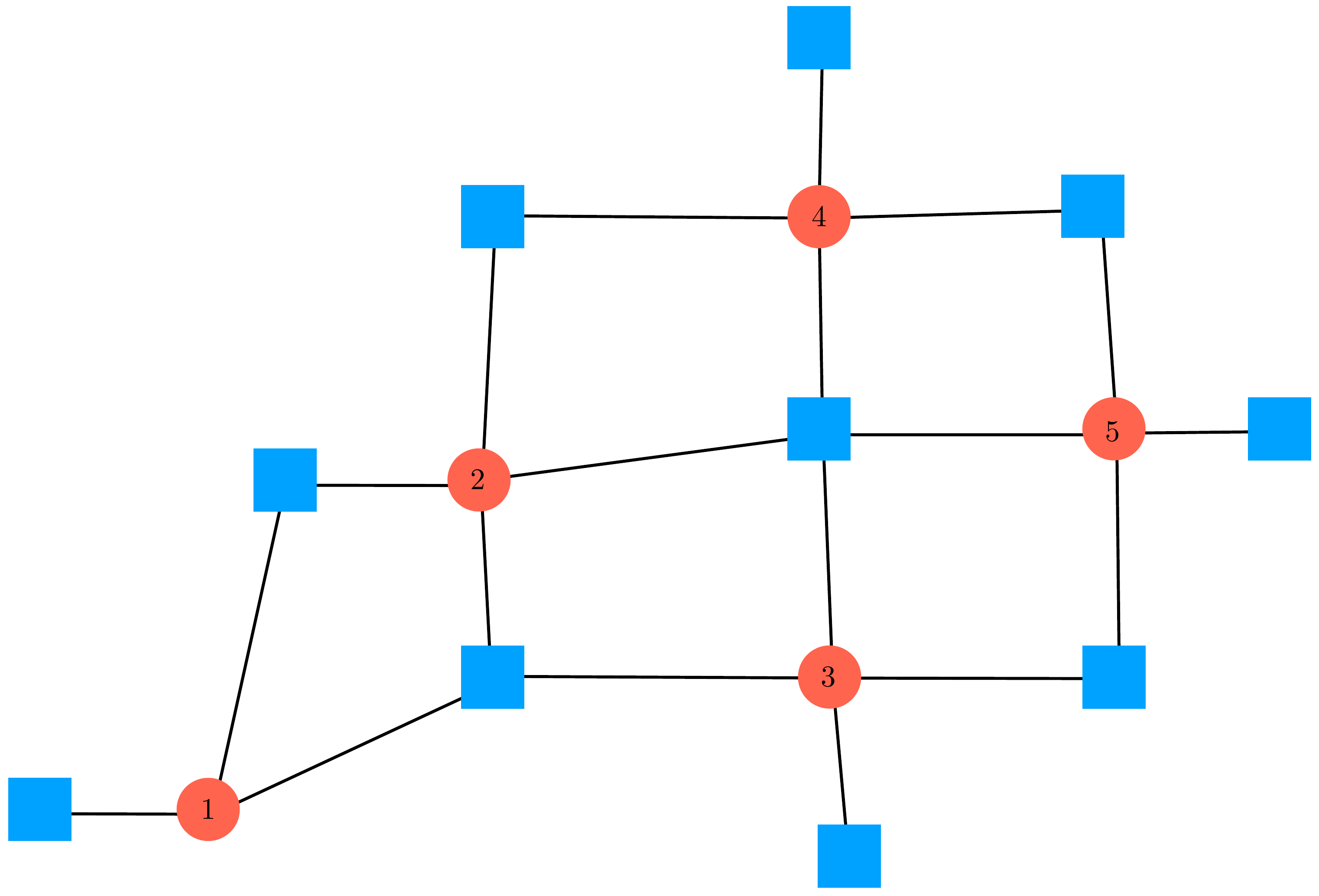}
\caption{An example  of  a factor graph.  Red circles denote vertices $\in V$,  blue squares denote factor vertices $\in F$, and black lines denote edges.   A Hamiltonian  which might  be associated with this factor graph is $H=\sigma_1^z + \sigma_3^x+\sigma_4^x+\sigma_5^y + \sigma_1^z \sigma^2_x  + \sigma_2^x\sigma_3^y + \sigma_2^x \sigma^y_4 + \sigma^z_3 \sigma^z_5 + \sigma_4^x \sigma_5^y + \sigma_1^z\sigma_2^z\sigma_3^z + \sigma_2^x\sigma_3^x\sigma_4^x\sigma_5^z$.}
\label{fig:factorgraph}
\end{figure}

In this section we derive Lieb-Robinson bounds for \emph{arbitrary} Hamiltonians of the form \begin{equation}
H = \sum_{S\subseteq \lbrace 1,\ldots,L\rbrace}  J_S  \tilde{H}_S,   \label{eq:arbitraryH}
\end{equation} 
where the  sum over $S$ includes all  possible  non-empty subsets of vertices, $J_S\ge0$, and $\tilde{H}_S$ is  an arbitrary unit norm Hermitian operator which acts non-trivially only on the Hilbert space $\bigotimes_{i\in S} \mathcal{H}_i$.   Observe that the requirement that the Hamiltonian is  $k$-local amounts to a requirement that $J_S=0$ if $|S|>k$.    To derive  our bound, we  review the useful notion of a factor graph \cite{montanari}.  A factor  graph $\tilde{G} = (V,F,E)$ consists of a vertex set $V=\lbrace 1,\ldots, L\rbrace$, a ``factor vertex set" $F\subset \mathbb{Z}_2^V$, which  consists of the possible $S$ in (\ref{eq:arbitraryH}), and an edge set $E \subset V\times F$ containing all edges obeying the rule \begin{equation}
e = (v,S) \left\lbrace \begin{array}{ll}  \in E &\ v\in S \\ \notin E  &\ v\notin S\end{array}\right..
\end{equation}  See Figure \ref{fig:factorgraph}.   
As we will see below, it is convenient to take the factor vertex  set $F$ to only include the sets $S$ for which $J_S > 0$,  together with all $S$ of size 1 (corresponding to single vertices), but this is not strictly necessary.   In this case, the factor vertices denote terms in $H$, and the edges from $F$ to $V$ simply denote which vertices each term ``acts" upon.   
We can define a natural adjacency matrix \begin{equation}
\mathcal{A}_{vS} = \left\lbrace \begin{array}{ll}  1 &\   (v,S) \in E \\ 0 &\ (v,S)\notin E \end{array}\right.,
\end{equation}
and we also define the matrix  \begin{equation}
\mathcal{J}_{SS^\prime} = \mdelta_{SS^\prime} J_S.
\end{equation}

Let us  now return to the question of operator growth.   Consider a pair of operators $A_S$ and $B_Q$ which act non-trivially only on the vertices in $S\in F$ and $Q\in F$, respectively.    Define \begin{equation}
C_{SQ}(t) \equiv  \sup_{A_S, B_Q} \frac{\lVert [A_S(t),B_Q]\rVert }{\lVert  A_S\rVert\lVert B_Q\rVert},
\end{equation}
with the supremum running over all possible Hermitian operators $A_S$ and $B_Q$ acting on the suitable Hilbert spaces.   Without loss of generality below, we take $t>0$ and $\epsilon>0$.   Our goal  is to find an inequality  governing the growth of $C_{SQ}(t)$.    Defining \begin{equation}
H_S \equiv  \sum_{S^\prime : S\cap S^\prime \ne \emptyset} J_{S^\prime} \tilde{H}_{S^\prime}.
\end{equation}  we observe that 
\begin{align}
\lVert [A_S(t+\epsilon),B_Q]\rVert &- \lVert [A_S(t),B_Q]\rVert  = \lVert [A_S(t),B_Q] - \mathrm{i}\epsilon  [[H,A_S(t)],B_Q]\rVert - \lVert [A_S(t),B_Q]\rVert   + \mathrm{O}\left(\epsilon^2\right)  \notag  \\
& \le \epsilon \lVert [[H,A_S],B_Q(-t)]\rVert + \mathrm{O}\left(\epsilon^2\right) = \epsilon \lVert [[H_S,A_S],B_Q(-t)]\rVert+ \mathrm{O}\left(\epsilon^2\right) \notag \\
&\le 2\epsilon  \lVert A_S\rVert \lVert [H_S(t),B_Q]\rVert+ \mathrm{O}\left(\epsilon^2\right)\end{align}
We thus obtain \begin{align}
C_{SQ}(t+\epsilon) &= \sup_{A_S,B_Q} \left[ \frac{\lVert [A_S(t),B_Q]\rVert }{\lVert  A_S\rVert\lVert B_Q\rVert} + 2\epsilon \frac{\lVert [H_S(t),B_Q]\rVert }{\lVert B_Q\rVert} \right]+ \mathrm{O}\left(\epsilon^2\right)  \notag  \\
&\le C_{SQ}(t) + 2\epsilon \sum_{S^\prime : S\cap S^\prime \ne \emptyset} J_{S^\prime} C_{S^\prime Q}(t)+ \mathrm{O}\left(\epsilon^2\right) \notag  \\
&\le C_{SQ}(t) + 2\epsilon \sum_{S^\prime} |S^\prime \cap S|  J_{S^\prime} C_{S^\prime Q}(t)+ \mathrm{O}\left(\epsilon^2\right),
\end{align}
which, upon taking $\epsilon  \rightarrow 0$, leads to the differential inequality \begin{equation}
\frac{\mathrm{d}C_{SQ}(t)}{\mathrm{d}t} \le 2   \mathcal{A}_{Sv} \mathcal{A}_{vS^{\prime\prime}}   \mathcal{J}_{S^{\prime\prime}S^\prime} C_{S^\prime Q}.  \label{eq:dCSQdt}
\end{equation}
where we have employed the Einstein summation convention  above.    In order to obtain (\ref{eq:dCSQdt}), we have used the  fact that the matrix $(\mathcal{A}^{\mathsf{T}}\mathcal{A} )_{SS^\prime} = |S\cap S^\prime|$ counts the number of vertices which the two sets $S$ and $S^\prime$ share in common.   Observe that both $C_{SQ}(0)$ and $\mathcal{A}^{\mathsf{T}}\mathcal{AJ}$ are non-negative matrices --  i.e., all components are non-negative.   We can  integrate (\ref{eq:dCSQdt}) to obtain \begin{equation}
C_{SQ}(t) \le \exp\left[ 2t \mathcal{A}^{\mathsf{T}}\mathcal{AJ}\right]_{SS^\prime} C_{S^\prime Q}(0) = \sum_{n=0}^{\infty} \frac{(2t)^n}{n!} \left( \mathcal{A}^{\mathsf{T}}\mathcal{AJ}\right)^n_{SS^\prime} C_{S^\prime Q}(0). \label{eq:CSQt}
\end{equation}
By definition, we know that \begin{equation}
C_{SQ}(0) \le   \left\lbrace \begin{array}{ll} 2 &\  S\cap Q \ne \emptyset \\  0 &\ S\cap Q = \emptyset \end{array}\right. \le 2  (\mathcal{A}^{\mathsf{T}}\mathcal{A} )_{SQ}.
\end{equation}

Suppose for simplicity that we are interested in the  evolution of operators that act only on a single vertex at time $t=0$.   In that case, the initial sets $S=\lbrace u\rbrace$ and $Q = \lbrace v\rbrace$, and we may write $\mathcal{A}^{\mathsf{T}}_{Su^\prime} Z_{u^\prime v^\prime} \mathcal{A}_{v^\prime Q} = Z_{uv}$ for any matrix $Z_{uv}$.   Using this identity, together with (\ref{eq:CSQt}), and denoting $C_{\lbrace u \rbrace \lbrace v\rbrace}(t) = C_{uv}(t)$, we obtain our most general Lieb-Robinson bound:\begin{equation}
C_{uv}(t) \le 2 \sum_{n=0}^{\infty} \frac{(2|t|)^n}{n!} \left( \mathcal{AJ} \mathcal{A}^{\mathsf{T}}\right)^n_{uv} =2 \exp\left[ 2|t|  \mathcal{AJ} \mathcal{A}^{\mathsf{T}} \right]_{uv} \label{eq:LRgen}
\end{equation}

Let us now focus on 2-local Hamiltonians with only $h_{uv}^{\alpha\beta}\ne 0$, as in the introductory section. Using (\ref{eq:maincdef}) we obtain \begin{align}
\left(\mathcal{AJ} \mathcal{A}^{\mathsf{T}}\right)_{uv} &=  \frac{1}{2}\sum_{u^\prime, v^\prime}  J_{\lbrace u^\prime v^\prime\rbrace } \mathcal{A}_{\lbrace u\rbrace (u^\prime,v^\prime)}\mathcal{A}^{\mathsf{T}}_{ (u^\prime,v^\prime)\lbrace v\rbrace} \notag \\
&\leq \frac{c}{2K} \sum_{u^\prime v^\prime \in E}\left(\mdelta_{uu^\prime} + \mdelta_{uv^\prime}\right)\left(\mdelta_{vu^\prime} +\mdelta_{vv^\prime} \right) = \frac{c}{K}(D+A)_{uv} .  \label{eq:LRgen2}
\end{align}
In the sum on the first line, we put a factor of $\frac{1}{2}$ to account for the double  counting of $u^\prime v^\prime$ and $v^\prime u^\prime$.   Combining (\ref{eq:LRgen}) and (\ref{eq:LRgen2})  we find  (\ref{eq:LR1st}).

\subsection{Fast Scrambling on ``Regular Hypergraphs"}
In the main text, we described the constraints on fast scrambling in 2-local models.   To generalize  this to $k$-local models, we can now consider the following simple argument.  Let us suppose that  the number of $J_S>0$ is given by $\mathcal{N}$.  For simplicity, let us assume that  \begin{equation}
J_S \le \frac{LJ_*}{\mathcal{N}},
\end{equation}
with $J_*>0$, though it is straightforward to generalize the argument.   Let $\tilde{1}_v = (1,\ldots,1)$. 
Since \begin{equation}
\left(\mathcal{JA}^{\mathsf{T}}\right)_{Sv} \frac{\tilde{1}_v}{L}  \le  \frac{J_*}{\mathcal{N}} |S| \le \frac{kJ_*}{\mathcal{N}}, \label{eq:JAT}
\end{equation} and thus \begin{equation}
 \left(\mathcal{AJA}^{\mathsf{T}}\right)_{uv} \frac{\tilde{1}_v}{L} \le \frac{kJ_*}{\mathcal{N}}  \sum_{S:  u\in S} 1 = \frac{kJ_*}{\mathcal{N}} \mathfrak{n}_u,
\end{equation}
where $\mathfrak{n}_u$ denotes the number of terms in $H$ that act non-trivially on $u$.    This generalizes the degree $k_u$ of the vertex in the 2-local case.     On a hypergraph, where $\mathfrak{n}_u \le m \mathcal{N}/L$ for every vertex (note $m\ge 1$),  we thus arrive at \begin{equation}
 \left(\mathcal{AJA}^{\mathsf{T}}\right)_{uv} \frac{\tilde{1}_v}{L} \le kJ_* m \frac{\tilde{1}_u}{L},
\end{equation}
and hence \begin{equation}
\sum_{u,v} \frac{C_{uv}(t)}{L^2} \le \frac{\mathrm{e}^{2kJ_* mt}}{L}.
\end{equation}
If both $k$ and $m$ are finite in  the limit $L\rightarrow \infty$, the exponent is finite and thus $C_{uv}$ only becomes O(1) after a time $t\sim \log L$, consistent with the fast scrambling conjecture.

\subsection{Towards the Traditional Bound}
We  now derive  (\ref{eq:LRdistmain}), along with a generalization for arbitrary Hamiltonians.   The observation is simply that we can tighten (\ref{eq:JAT}) to \begin{equation}
 \left(\mathcal{AJA}^{\mathsf{T}}\right)_{uv} \frac{1_v}{L} \le \frac{1_u}{L} \times   \left\lbrace \begin{array}{ll}  (kJ_* m)^n &\   n\ge d_{uv} \\  0 &\ n<d_{uv}  \end{array}\right.,
\end{equation}   
where $d_{uv}$, the distance between $u$ and $v$, is the straightforward generalization of distance to the factor graph: half of the number of edges in the factor graph which must be traversed.   From (\ref{eq:LRgen}): \begin{equation}
C_{uv}(t) \le 2\sum_{n=d_{uv}} \frac{(2t)^n}{n!} \left( \mathcal{AJ} \mathcal{A}^{\mathsf{T}}\right)^n_{uv} < 2\mathrm{e}^{-d_{uv}}\sum_{n=d_{uv}}\frac{(2\mathrm{e}t)^n}{n!} (kJ_* m)^n < 2\exp\left[2\mathrm{e}kJ_*mt    - d_{uv} \right].
\end{equation}
Using (\ref{eq:LRgen2}), we obtain (\ref{eq:LRdistmain}).   More generally, we see that if $m$ (as defined in  the previous subsection) and $k$ are finite, fast scrambling is only possible on hypergraphs with diameter $\lesssim \log L$.

\section{Sachdev-Ye-Kitaev Models on a Graph}
\label{sec:SYK}
In this section, we describe the generalized Sachdev-Ye-Kitaev model \cite{gu1609, gu1702} on a random graph.  Let $\chi^v_i$ denote Majorana fermion $i$ on  vertex $v\in V$:  $\lbrace \chi^v_i, \chi^u_j\rbrace = \mdelta_{ij}\mdelta^{uv}$, and $i\in \lbrace 1,2,\ldots,M\rbrace$, and consider the Hamiltonian  
\begin{equation}
H =\mathrm{i}^{\frac{q}{2}}  \sum_u  \sum_{i_1< \ldots < i_q} J_{i_1,\ldots,i_q}^u \chi^u_{i_1} \ldots \chi^u_{i_q} +\mathrm{i}^{\frac{q}{2}} \sum_{uv  \in E} \sum_{\substack{i_1<\ldots<i_{q/2}\\j_1<\ldots<j_{q/2}}} {J}^{uv}_{i_1,\ldots,i_{q/2} j_1,\ldots,j_{q/2}} \chi_{i_1}^u \ldots \chi_{i_{q/2}}^u \chi_{j_1}^v\ldots \chi_{j_{q/2}}^v .
\end{equation}
where we assume Gaussian random variables $J^u_{i_1,\ldots,i_q}$ and ${J}^{uv}_{i_1,\ldots,i_{q/2} j_1,\ldots,j_{q/2}}$ are independent and mean zero with the following variants:
\begin{equation}
\mathbb{E}\left[\left(J^u_{i_1,\ldots,i_q} \right)^2 \right] = \frac{(q-1)!}{M^{q-1}} \frac{2^{q-1}}{q}  \left(1 - \frac{b k_u}{2}\right) J^2, \;\;\;\; 
\mathbb{E}\left[\left({J}^{uv}_{i_1,\ldots,i_{q/2} j_1,\ldots,j_{q/2}} \right)^2 \right] = \frac{(q/2)!^2}{ q M^{q-1}}  \frac{2^{q-1}}{q} b J^2.
\end{equation}
The parameter $0\leq b\leq \frac{2}{k_{\text{max}}}$ sets the relative coupling strength on the links;  ${J}$ sets the effective coupling constant.    

\subsection{Large $q$ Limit  and Saddle Point}

At large $M$, one solves for the correlation functions of the fermions by computing an effective action for the two-point Euclidean time Green's function \begin{equation}
G_u(\tau_1,\tau_2) = \frac{1}{M}\sum_{i=1}^M \langle \chi^u_i(\tau_1)\chi_i^u(\tau_2)\rangle,
\end{equation} and an analogous self-energy $\Sigma_u(\tau_1,\tau_2)$.  The result is 
\begin{align}
S_{\eff}=\sum_{u,v} \left[ -\log \Pf \left( \partial_\tau - \Sigma_{u}  \right)  \delta_{uv} + \frac{1}{2} \int \mathrm{d}\tau_1 \mathrm{d}\tau_2 \left(
\Sigma_u G_u \delta_{uv} - \frac{J^2}{q} ( G_u^q \delta_{uv} -\frac{1}{2} b \Lambda_{uv} G_u^{q/2} G_v^{q/2}  )
\right) \right],  \label{eq:SYKSeff}
\end{align}
where the matrix $\Lambda=D-A$  is the graph Laplacian, which generalizes a discretized $-\nabla^2$ to a general graph.   $\Lambda$ encodes all spatial dynamics of the effective action. 
Because $\Lambda$ always has a null vector:  \begin{equation}
1_v = \frac{1}{\sqrt{L}}(1,1,\ldots, 1),
\end{equation}
the above action admits a simple saddle point, where $G_v = G_*$ and $\Sigma_v=\Sigma_*$ do not depend on vertex $v$.  In the large $q$ limit,  
we expand $G_*$ and $\Sigma_*$ to $\mathrm{O}(1/q)$ for this saddle point: \cite{maldacena2016remarks} \begin{subequations}\begin{align}
G_*(\tau) = \frac{1}{2} \sgn (\tau) \left( 1+ \frac{g_*(\tau)}{q} \right), \quad
 \Sigma_*(\tau) = \frac{{\cal J}^2}{q} \mathrm{e}^{g_*(\tau)},
\end{align}
\end{subequations}
where the saddle point solution $g_*(\tau)$  is given by
\begin{align}
\mathrm{e}^{g_*(\tau)}= \left[ \frac{\cos \frac{\mpi \eta}{2} }{ \cos \mpi \eta \left( \frac{1}{2} - \frac{|t|}{\beta} \right) }  \right]^2, \quad \beta \calJ = \frac{\mpi \eta}{\cos \frac{\mpi \eta}{2}}
\end{align}
where $\eta$ is a parameter determined by the coupling constant:  in the high temperature limit $\beta \calJ \ll 1$, $\eta\approx \frac{\beta \calJ}{\pi}$;  in the low temperature limit  $\beta\calJ \gg 1$, $\eta\approx 1- \frac{2}{\beta \calJ}$. 

In the discussion that follows, we will assume that $q$ is large, but not the largest parameter in the problem.   Namely, we will take  $M\gg L \gg q$.   We assume that the latter inequality may be safely taken, but have not explicitly checked this assumption.

\subsection{Out-Of-Time-Ordered Correlators}
By expanding around this saddle point, we can compute the connected piece of the averaged, regularized OTOC
\begin{equation}
\sum_{i,j=1}^M \frac{ \mathrm{tr} \left(y \chi_i^u (0) y \chi^v_j(t_1)  y \chi^u_i(0) y \chi_j^v(t_2) \right)_{\text{conn.}} }{M^2} \equiv \mathcal{F}_{uv}(t_1,t_2),
\end{equation}
where $y = \rho_\beta^{1/4}$, with $\rho_\beta$ the thermal density matrix at inverse temperature $\beta$.
Because of the combined the large $M$ limit and large $q$ limit, we can describe the exponentially growing behavior of $\mathcal{F}_{uv}(t_1,t_2)$ at all temperatures in this model.    Indeed, following \cite{kitaevunpublished,maldacena2016remarks}, we find that $\mathcal{F}_{uv}$ obeys the following linear equation in the exponential growth regime: 
\begin{align}
\mathcal{F}_{uv}(t_1,t_2) = \int \mathrm{d}t_3 \mathrm{d}t_4 \; K^{\mathrm{R}}_{uw}(t_1,t_2;t_3,t_4) \mathcal{F}_{wv} (t_3,t_4), \label{eqn: kinetic equation}
\end{align}
where $K_{uv}^R$ is a retarded kernel with `spatial' dynamics:
 \begin{equation}
K^{\mathrm{R}}_{uv}(t_1,t_2;t_3,t_4) = \frac{2\mpi^2 v^2 \mathrm{\Theta}(t_{13})\mathrm{\Theta}(t_{24})}{\beta^2 \cosh^2 \left(\frac{\mpi \eta}{\beta}t_{34}\right)} S_{uv},
\end{equation}
and\footnote{See \cite{gu1609,gu1702} for similar discussions and diagrams for spatially dependent kernels in generalized SYK models on lattices.}
\begin{equation}
S_{uv} = \mdelta_{uv} - \frac{b}{q-1} \Lambda_{uv}.
\end{equation}
The exponentially growing ansatz has the following form:
\begin{equation}
\mathcal{F}_{uv}(t_1,t_2) = \mathrm{e}^{\lambda_{\rm L} \frac{t_1+t_2}{2}} f_{uv}(t_{12}).
\end{equation}
Because the vertex dependence in $K^{\mathrm{R}}_{uv}$ comes entirely through $S_{uv}$, we can solve for the spatial dynamics by studying (\ref{eqn: kinetic equation}) for each individual eigenvector $\phi$ of $S$ with eigenvalue $s$.  Suitable combinations of the solution to (\ref{eqn: kinetic equation}) for these eigenvectors can be used to construct $\mathcal{F}_{uv}$ for any initial conditions.  Using the explicit form of the retarded kernel, applying a pair of derivatives,  $\partial_1 \partial_2 $, to (\ref{eqn: kinetic equation}), and denoting $t_{12}= \frac{\beta u}{\mpi v}$, we find:
\begin{align}
\frac{\lambda_{\rm L}^2 \beta^2 }{4 \mpi^2 \eta^2 } \cdot f_s(u) = \left(\partial_u^2+ \frac{2s}{\cosh^2u } \right) f_s(u)
\end{align}
This equation is the Schr\"odinger equation in a cosh-potential in one spatial dimension.  The following exact bound state solution is known, together with a corresponding eigenvalue parametrized by $a$:
\begin{align}
f_s(u) \propto \frac{1}{\cosh^{a} u } \quad \Rightarrow \quad \frac{\lambda_{\rm L}^2 \beta^2 }{4 \pi^2 \eta^2 } = a^2, \quad 2s=a (a+1) .
\end{align}
Importantly, we have now fixed the Lyapunov exponent $\lambda_{\mathrm{L}}$.   Focusing for simplicity on the OTOC $\mathcal{F}_{uv}(t) = \mathcal{F}_{uv}(t,t)$, 
we conclude that 
\begin{equation}
\mathcal{F}_{uv}(t)  = \frac{1}{M}\sum_\lambda  C_\lambda \mathrm{e}^{\lambda_{\mathrm{L}}(s)t} \phi^\lambda_u \phi^\lambda_v,  
\label{eq:FuvC}
\end{equation}
where $C_\lambda$ are undetermined constants, and $\phi^\lambda_v$ denote eigenvectors of $\Lambda$, and therefore $S$:  
\begin{equation}
\sum_{v}
S_{uv}\phi^\lambda_v =  s\phi^\lambda_u = \left(1-\frac{b\lambda}{q-1}\right)\phi^\lambda_u.
\end{equation}
In the large $q$ limit, the eigenvalues of $S_{uv}$, $s$, are close to $1$.   Therefore
\begin{align}
\lambda_{\rm L}= \frac{2\mpi}{\beta} \eta a 
\approx \frac{2\mpi}{\beta} \eta \left(1-\frac{2b\lambda}{3(q-1)}\right).  \label{eq:lambdaLhighT}
\end{align}

It remains to fix $C_\lambda$ to solve the problem.  Unfortunately, this is highly non-trivial in general, and depends on complicated details of the early time physics in the SYK model, e.g. on  the dynamics at early time $0<t<\beta$ where (\ref{eqn: kinetic equation}) doesn't apply.  
 We do understand, however, 
the late  time dynamics in a number of important  limits, which we elucidate below. 
 

\subsection{High Temperatures}

First, we describe the high temperature limit 
$\beta J \ll 1$.  In this regime, $\eta\approx \frac{\beta \calJ}{\mpi}$ and $ \lambda_{\rm L}=2J\left(1-\frac{2b\lambda}{3(q-1)}\right)$.
It is reasonable to expect that $\mathcal{F}_{uv}(t)$ is approximately local at early times $0<t<\beta$.    To zeroth order, we then anticipate that (\ref{eq:FuvC}) can be applied at $t=0$, and that \begin{equation}
\sum_\lambda C_\lambda \phi^\lambda_u \phi^\lambda_v \propto \mdelta_{uv}.
\end{equation}
Since the set of $\phi_u^\lambda$ form an orthonormal basis, we conclude that $C_\lambda$ is a constant, independent of $\lambda$. 
This implies that
 \begin{equation}
\mathcal{F}_{uv}(t)\propto \frac{1}{M} \exp\left[2 Jt \left(1-\frac{2b}{3(q-1)} \Lambda\right)\right]_{uv}.  \label{eq:FuvhighT}
\end{equation}
We can interpret this result in a simple way:  $\mathcal{F}_{uv}(t)$ is proportional to a concentration of ``infected random walking individuals" located on vertex $u$ at time $t$, given that the only  infected individuals were located on vertex $v$ at time $t=0$.   The assumptions we make are that infected individuals grow at a constant rate of $2J$, and that infected individuals perform a random walk on $G$:  traversing any given edge at a constant rate of $\frac{4bJ}{3(q-1)}$.   This connection between the growth of chaos and the spread of infections has been observed for some time, and is also visible in the RUC.

\subsection{Graphs with Finite Spectral Gap}

Next, we relax the high temperature assumption, but require that the graph Laplacian $\Lambda$ has a finite spectral gap $\gamma>0$. 
The gap of $\Lambda$ implies a gap of Lyapunov spectrum, i.e.
\begin{equation}
\mathcal{F}_{uv}(t)  = \frac{1}{M} \left(  C_0 \mathrm{e}^{\frac{2\mpi}{\beta} \eta t} 1_u 1_v
+
C_1 \mathrm{e}^{\frac{2\mpi}{\beta} \eta (1-\frac{2b}{3(q-1)} \gamma) t } \phi^\gamma_u \phi^\gamma_v + \ldots
\right)
\end{equation}
In the long time limit, e.g. $t\gg q\beta$, the subleading terms are negligible comparing to the first term, related to the null vector $1_u= \frac{1}{\sqrt{L}}(1,1,\ldots,1)$.\footnote{Technically, we also require $ML = N \gg \mathrm{e}^q$ for this long time limit to be sensible.}
 Therefore we have:
\begin{equation}
\mathcal{F}_{uv}(t) \approx C_0 1_u 1_v \frac{\mathrm{e}^{\frac{2\mpi}{\beta} \eta  t}}{M} = C_0  \frac{\mathrm{e}^{\frac{2\mpi}{\beta} \eta  t}}{N}
\end{equation}
In the last step we used $N=ML$.   
 The constant
$C_0$ is set by the overlap of the initial condition with the spatially uniform component of $\calF_{uv}$, before the exponential growing regime, and we therefore expect $C_0$ to be an O(1)  number.   Thus, for the graphs with finite spectral gap, we obtain
\begin{equation}
t_* = \frac{\beta }{2\pi \eta}  \log N 
\label{eqn:tsr}
\end{equation}
at leading order in $N$, independently of any details of the graph structure.   

As we are more interested in the spatial dynamics on the graph than in the dynamics of a single-site SYK model, we wish to take the limits $M,L\rightarrow \infty$ in such a way that $\frac{\log M}{L}\rightarrow 0$.   In this case, it is necessary for the spectral gap $\gamma$ to remain finite even in the thermodynamic limit $L\rightarrow \infty$, in order to obtain (\ref{eqn:tsr}).   Rather remarkably, it is a famous result in graph theory that $\gamma$ is strictly finite on any graph where the `perimeter' of any subset of vertices scales proportionally to the number of vertices: see Appendix \ref{app:graphs}.   Therefore, on a typical sparse and locally treelike graph, the SYK model described above is just as chaotic as it would be on a fully connected graph. This serves as an explicit example of a chaotic quantum system where some amount of sparsity to the connectivity graph does not affect the time scales of quantum information loss and operator growth.   


\subsection{Low Temperatures}
Another commonly discussed solvable limit is the low temperature limit $M\gg \beta J \gg 1$.  In  this limit, we do not need to assume that $q$ is large.
In this limit, the saddle point equation can be approximately solved by a conformal ansatz, where $K^{\mathrm{R}}_{uv}$ takes the following form (for simplicity, we set $q=4$ for this subsection):
\begin{equation}
K^{\mathrm{R}}_{uv}(t_1,t_2;t_3,t_4) = 
\frac{3\mpi \mathrm{\Theta}(t_{13})\mathrm{\Theta}(t_{24})}
{\beta^2 \cosh \left(\frac{\mpi }{\beta}t_{34}\right) \left( \sinh \left(\frac{\mpi }{\beta}t_{13} \right) \sinh \left(\frac{\mpi }{\beta}t_{24}\right) \right)^{\frac{1}{2}}} 
S_{uv},
\end{equation}
Similarly to the large $q$ discussion, $K^{\mathrm{R}}_{uv}$ can be diagonalized by the eigenvectors of $S_{uv}$, and eigenfunctions of the temporal part \cite{maldacena2016remarks, kitaevunpublished}, which determines the Lyapunov exponent through the following equation
\begin{align}
\frac{3}{1+\frac{\beta}{\mpi} \lambda_{\mathrm{L}}} \left( 1 - \frac{b\lambda}{3} \right)=1.
\end{align}
We  hence obtain \begin{align}
 \lambda_{\mathrm{L}} = \left( 1 -\frac{b\lambda}{2} \right) \frac{2\mpi}{\beta}. \label{eq:blambdaLlowT}
\end{align}
Therefore the gap $\gamma$ in $\Lambda$ also indicates a gap in the Lyapunov spectrum for the $q=4$ model at low temperature (a similar result also applies to arbitrary $q\ge4$). Following the same argument as before, we can ignore the subleading terms and only focus on the null vector $1_u$ and leading exponent $\lambda_{\rm L}= \frac{2\pi}{\beta}$, which saturates the chaos bound  \cite{stanfordbound}: 
\begin{equation}
{\mathcal F}_{uv} (t)\propto \frac{\mathrm{e}^{\frac{2\mpi}{ \beta} t}}{ML}.
\end{equation}
independently of the graph structure.  


We can gain further intuition about the spatial dynamics by using an alternative treatment of the leading exponent in the low temperature limit. 
Following \cite{gu1609, gu1702}, 
we describe the contribution to the leading exponent  by the dynamics of reparametrization modes on every vertex.  The generalization of these works is straightforward but as the computation is rather technical we do not write it explicitly.  
We find that the OTOC becomes
\begin{align}
{\mathcal F}_{uv} (t) \propto \frac{1}{M} \left( \frac{\alpha}{\beta {J}}  + \frac{b}{3} \Lambda  \right)_{uv}^{-1} \mathrm{e}^{\frac{2\mpi}{\beta} t}\equiv \frac{\mathrm{e}^{\frac{2\mpi}{\beta} t}}{M} Z_{uv}^{-1}.
 \label{eq:FuvlowT}
\end{align}
where $\alpha$  is a numerical constant. 
The identity matrix is implicitly assumed to be multiplying constants in the expression above.   
 Note that this calculation assumes that $\lambda_{\mathrm{L}} \approx \frac{2\mpi}{\beta}$ for all modes; we  see from (\ref{eq:blambdaLlowT}) that this is analogous to approximating $b\rightarrow 0$.


At leading order in the small parameter ${1}/{\beta J}$, we reproduce the previous result $Z_{uv}^{-1} \approx \frac{\beta J}{\alpha L}$ independent of indices $u$ and $v$. Regarding the higher order effects, 
we can interpret $Z_{uv}^{-1}$ in terms of a simple statistical problem: at vertex $u$, we release one random walker per unit time onto the graph.   At any given instant in time, a random walker may walk along an edge of the graph to a neighboring vertex -- this occurs with rate $\frac{b}{3}$.   Finally, the random walkers die with rate constant $\frac{\alpha}{\beta J}$.    Let $n_v$ be the expected number of random walkers on vertex $v$.  In steady state, the rate of incoming walkers equals the rate  of outgoing/dying walkers:  \begin{equation}
\mdelta_{uv} + \frac{b}{3} \sum_{wv\in E} n_w = \left(\frac{\alpha}{\beta J}+\frac{b}{3}k_v\right)n_v. \label{eq:randomwalk}
\end{equation}
We conclude that $n_v = Z^{-1}_{uv} \ge 0$.   Interestingly, this does not have the interpretation of a diffusing random  walker.   The meaning of the different mechanisms for   the spread of chaos at low vs. high temperature is not clear to us, although its origins can be straightforwardly understood: the mechanisms responsible for chaos at high \cite{stanford1802} and low  \cite{maldacena2016remarks} temperatures are very different.  See also the recent discussion in \cite{swingle1805}, which argues that the high temperature behavior is more generic at finite $M$.

\section{Random Unitary Circuit Model}
\label{app:RUC}


\subsection{Mapping OTOCs to a Classical Stochastic Process}

In this subsection, we review the stochastic process which the computation of OTOCs in the $m$-local RUC maps on to.  These results were all found previously in \cite{nahum}.  

First, consider a growing operator $\mathcal{O}_v(\mathrm{\Delta} t) = U\mathcal{O}_v U^\dagger$, with $\mathcal{O}_v$ a local operator which acts non-trivially only on vertex $v$.  Recall that $\mathrm{\Delta}t=1/L$ in order to recover extensive quantum dynamics.  For  simplicity, let us assume that $U$ will act on a subset $A$ of $m$ vertices:   namely, $U$ is a random $2^{Mm}\times 2^{Mm}$ unitary matrix.    $U\mathcal{O}_v U^\dagger = \mathcal{O}_v$ if $v\notin A$.  If  $v\in A$, after averaging over all $U$ with  uniform measure, $U\mathcal{O}_v U^\dagger$  will be an equal superposition of all non-trivial $2^{2Mm}-1$ Hermitian operators acting on the  subset $A$.   Observe that the fraction of these operators which act non-trivially  on $\ell  \le m$ vertices is given by \begin{equation}
p_\ell  =  \frac{(2^{2M}-1)^\ell}{2^{2Mm}-1}  \left(\begin{array}{c} m \\ \ell \end{array}\right).
\end{equation}
If we fix  vertices $u$ and $v$, and average over all $U$s acting on the subset $A$, and also average over local operators $\mathcal{O}_u$ and $\mathcal{O}_v$, we find \begin{equation}
\mathbb{E}\left[ \left\lVert \mathcal{O}_u, U \mathcal{O}_vU^\dagger \right\rVert^2   \right] \propto \left\lbrace \begin{array}{ll} \displaystyle  \sum_{\ell=1}^m \frac{\ell}{m}p_\ell &\ \lbrace u, v\rbrace \subset A \\ 0 &\ \text{otherwise} \end{array}\right..
\end{equation}
The proportionality constant is related to the  normalization of the $\mathcal{O}_v$ and is not needed for our purposes.   The right hand side should be  interpreted as the weight in the operator $U\mathcal{O}_v U^\dagger$ which act non-trivially on $u$.

At later time steps, the above procedure generalizes straightforwardly.   For an operator $\mathcal{A} =  \sum \bigotimes \mathcal{O}_i $ which consists of  a large number of complicated terms,  $U \mathcal{A} U^\dagger$ can be evaluated term-by-term.   In  each term $\bigotimes \mathcal{O}_v$,  if there exists a vertex $v\in A$  for which $\mathcal{O}_v$ is not the identity,  then we replace $\bigotimes_{u\in A} \mathcal{O}_u$ with a random sum of all possible $2^{2Mm}-1$ Hermitian operators with coefficients whose squares sum to one.

This leads to the following observation: after averaging over all possible circuits, we can compute the proportion of the growing operator $U(t) \mathcal{O}_v U(t)^\dagger$ which acts non-trivially on the subset $S\subset V$ by mapping onto the following stochastic process.   Let $n_u \in  \lbrace 0,1\rbrace$  denote whether a vertex is ``infected" or not; at time $t=0$, $n_v=1$ and all other $n_u=0$ for $u\ne  v$.   At each time step,  pick a  random allowed subset $A\subset V$ of $m$  vertices.   If $\sum_{u\in A} n_u = 0$, then do nothing;  otherwise -- regardless of the microscopic state -- with probability $p_\ell$,  set $\ell$ vertices in $A$, chosen uniformly at random, to be infected, and the remaining  $m-\ell$ vertices to be  uninfected.

In the limit $M\rightarrow \infty$,  $p_m=1$.  The infection only grows.   This  is why, when $m=2$, the RUC maps onto a discrete time analogue of the SI epidemic model.   This is the example which we focused on in the main text.

\subsection{2-Local Operator Dynamics: Mean Field Methods and their Breakdown}
\label{sec:RUC2}
As we saw at infinite $M$ in the main text, the RUC wih 2-local dynamics maps on to the SI epidemic model which has super-exponential infection growth on heterogeneous networks.  In this section, we will show that this effect is an artifact of the $M\rightarrow \infty$ limit for 2-local RUCs.   The following section will give an example of super-exponential operator growth with a 3-local RUC.

For now, we follow the literature on epidemics on complex networks \cite{vespignaniRMP} and use a mean field approximation to solve for the dynamics of the stochastic process of the previous section with $m=2$.     The key approximation is a closed set of equations for $P_v \equiv \mathbb{P}[n_v=1]$:  \begin{equation}
\frac{\mathrm{d}P_v}{\mathrm{d}t} = \frac{2}{K} \sum_{u\sim v} \left[(1-p_1)(1-P_v)P_u - p_1 P_v \right].
\end{equation}
The first term corresponds to the rate at which the vertex $v$ becomes infected:  this occurs with probability $1-p_1$ (as  we do not care  whether its neighbor gets uninfected in the process!), and requires the neighbor to be infected while $v$ should be uninfected.   The latter term corresponds to the rate at  which a vertex uninfects itself, which occurs at rate $2p_1/K$ per edge, regardless of the state of the neighbor.


At early times, we may write the  growth equation as \begin{equation}
\frac{\mathrm{d}P_k}{\mathrm{d}t}\approx \frac{2}{K} \left[(1-p_1)A_{vu}  - p_1 D_{vu}\right] P_u .
\end{equation}
The growth rate of the  epidemic is thus related to the spectrum of a particular linear combination of adjacency and degree matrices.  In order to make further progress, we resort to a further ``degree-based" mean-field description \cite{vespignaniRMP}. Let $P_k$ denote the probability that $n_v=1$ for a vertex $v$ with degree $k$, and let $\rho_k$ be the probability that a randomly chosen vertex in the graph has degree $k$.  We define \begin{equation}
\theta = \sum_k \frac{k\rho_k}{K}  P_k,  \label{eq:thetaPk}
\end{equation}
which is the probability that a randomly chosen edge points to a node which is infected.   The mean-field approximation  is that each node effectively sees each neighbor infected with probability $\theta$.  With this approximation, $P_k$ obeys the closed differential equation \begin{equation}
\frac{\mathrm{d}P_k}{\mathrm{d}t} = \frac{2k}{K} (1-p_1) (1-P_k)\theta - \frac{2k}{K} p_1P_k.  \label{eq:dPkdt}
\end{equation}
The overall factor of $2k/K$ arises because this is the rate at which a random edge which connects to a vertex of  degree $k$ is chosen.   The first term  in (\ref{eq:dPkdt}) counts the rate at which the vertex is uninfected, and an edge is chosen between the given vertex and one of its neighbors:  the final factor of $1-p$ counts the probability that the  infection spreads to the vertex.   The second term counts the probability that a random edge is chosen, and this causes the infection to decay from the central vertex.

At early times, we expect that $P_k \sim \mathrm{e}^{\lambda_{\mathrm{L}} t}$: namely, the largest eigenvalue of the linearized equation of motion dominates growth and sets a scrambling time, as defined by OTOCs.   This eigenvalue equation  gives that \begin{equation}
\left(K\lambda_{\mathrm{L}} +  2kp_1\right) P_k = 2k(1-p_1)\theta  \label{eq:thetaPk2}
\end{equation}   
when $P_k \rightarrow 0$.  Combining  (\ref{eq:thetaPk}) and (\ref{eq:thetaPk2}), we find that \begin{equation}
1 = \sum_k \frac{k\rho_k}{K} \frac{2k(1-p_1)}{K\lambda_{\mathrm{L}} + 2kp_1}.  \label{eq:thetafin}
\end{equation}

Let us begin with some formal bounds on  $\lambda_{\mathrm{L}} $.  A lower bound on $\lambda_{\mathrm{L}} $  can be  found from (\ref{eq:thetaPk2}) by  applying Jensen's inequality on  the convex function $x^2/(x+1)$ (for $x>0$):  \begin{equation}
1>   \frac{2K(1-p_1)}{K\lambda_{\mathrm{L}} + 2p_1K}.
\end{equation}
Hence \begin{equation}
\lambda_{\mathrm{L}}  > 2(1-2p_1).
\end{equation}
A lower bound can be found by applying  Jensen's inequality to the concave function $x/(x+1)$, assuming the probability distribution on degrees $k\rho_k/K$, instead of $\rho_k$:   \begin{equation}
\lambda_{\mathrm{L}}  < 2(1-2p_1) \frac{1}{K^2}\sum_k \rho_k k^2.
\end{equation}
This latter probability upper bound is proportional to  the growth rate of an SI epidemic on a heterogeneous graph \cite{vespignani}: $\frac{1}{K^2}\sum \rho_k k^2$.      Unsurprisingly, infections spread at a reduced rate at finite $M$.   This accordingly increases the time for the operator to grow on a regular graph by the factor $(1-2p_1)^{-1}$.     As noted in \cite{nahum}, this is  analogous to the emergence of a butterfly velocity \cite{localized}, which plays the role of  an effective Lieb-Robinson velocity, accessible in correlation functions \cite{lrbutterfly}.

In general, we expect that the upper bound above is better for heterogeneous graphs with a finite variance of the degree distribution, as  it is known that epidemics do spread faster on heterogeneous networks, and our model reduces to the SI model as $p_1\rightarrow 0$.   However, if $\sum_k \rho_k  k^2$ diverges, as it can on scale free graphs  with $\rho_k \propto K^{\nu-1}k^{-\nu}$, with $2<\nu \le 3$, then this  upper bound is  lousy.   We can estimate the value of $\lambda_{\mathrm{L}}$ on these scale free graphs by observing that \begin{align}
1 &\propto \int\limits_1^{K\lambda_{\mathrm{L}}/2p_1} \frac{\mathrm{d}k}{k^\nu} K^{\nu-1} \frac{2k^2(1-p_1)}{K^2\lambda_{\mathrm{L}}} + \int\limits^\infty_{K\lambda_{\mathrm{L}}/2p_1} \frac{\mathrm{d}k}{k^\nu}  K^{\nu-1}  \frac{k(1-p_1)}{Kp_1} \propto   \frac{1-p_1}{p_1} \left(\frac{\lambda_{\mathrm{L}}}{2p_1}\right)^{2-\nu},
\end{align}
which gives us that as $p_1\rightarrow  0$ \begin{equation}
\lambda_{\mathrm{L}}  \propto p_1^{-\frac{3-\nu}{\nu-2}} .
\end{equation}
This suggests that for any finite $p_1$, the ``epidemic spreading" dynamics of the RUC is fundamentally different from the SI model:  in particular,  $\lambda_{\mathrm{L}}$ remains finite.

Indeed, for 2-local dynamics, we are not able to construct any explicit examples where we can prove that $\lambda_{\mathrm{L}} = \infty$.    One way to try is to consider the ``star graph"  (the interaction graph of (\ref{eq:sigmazstar})).   Here (\ref{eq:thetafin}) becomes \begin{equation}
1 \approx   \frac{1-p_1}{2\lambda_{\mathrm{L}} + 2p_1} + \frac{ L(1-p_1)}{ 2\lambda_{\mathrm{L}}+2Lp_1 }.
\end{equation}   This equation is approximately solved by \begin{equation}
\lambda_{\mathrm{L}} \approx L(1-2p_1);
\end{equation} namely, the mean-field description implies that $\lambda_{\mathrm{L}}$ diverges in the thermodynamic limit $L\rightarrow \infty$.   However, on the star graph, we may also give  a more explicit construction of the dynamics.   For simplicity, let us consider dynamics where the infection starts on the central node, 1.  At time steps $\propto 1/L$, we act on the central node (and another) with a random unitary;  if the central node is infected,  at  each time step there is a finite probability $p_1$ that this node is ``uninfected".   With high probability, at $t\propto  1/(p_1L)$, the central vertex becomes uninfected.   If  $Q\propto p_1^{-1}$  of the other vertices were infected in this time frame, the time it takes for the central vertex to be reinfected is $\approx Q^{-1} \propto p_1$.   So we can estimate that the time it takes to infect a finite fraction of vertices as \begin{equation}
t\approx p_1 + \frac{p_1}{2}+ \frac{p_1}{3} +\cdots \sim p_1 \log L.
\end{equation}
This gives us a finite Lyapunov rate:  \begin{equation}
\lambda_{\mathrm{L}} \approx \frac{1}{p_1}.
\end{equation}
We have not mathematically shown that any 2-local RUC on any graph has a  larger Lyapunov exponent than this.  It is possible that the mean field treatment  of the dynamics on scale free graphs  is also inaccurate.

\subsection{A 3-Local Model with $\lambda_{\mathrm{L}}=\infty$}
\label{sec:RUC3}
\begin{figure}
 \centering
 \includegraphics[width=2in]{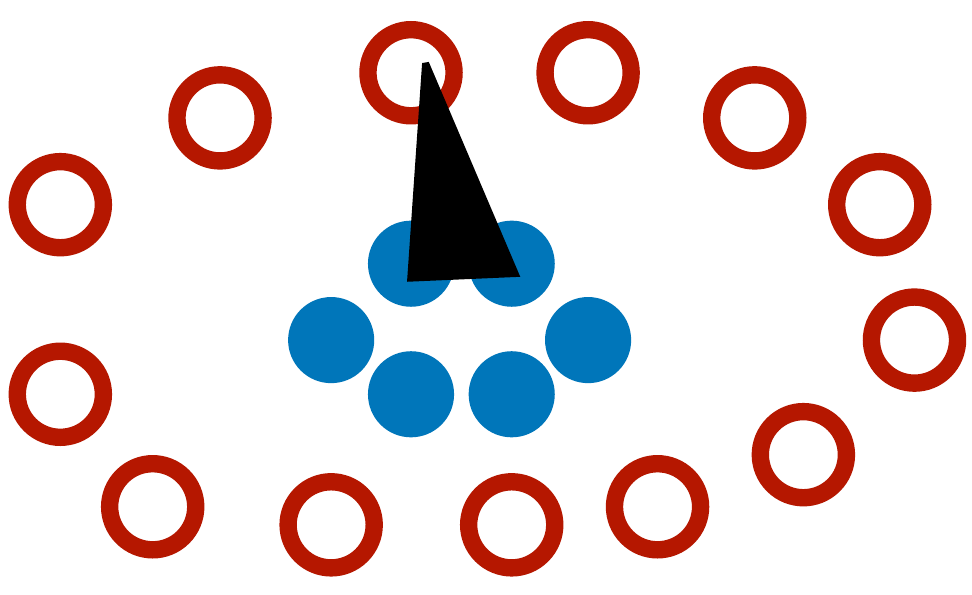}
 \caption{A 3-local RUC on a generalized star graph consisting of $R$ inner nodes (solid circles) and $N$ outer nodes (hollow circles).  The black triangle denotes a sample triplet of vertices on which a unitary can act.}
 \label{fig:3star}
 \end{figure}

One may ask -- is the observation that $\lambda_{\mathrm{L}} < \infty$ when $M$ is finite a general feature of the RUC, or is it an artifact of the 2-local dynamics above.  We now show that it is the latter.  Consider a 3-local RUC on a generalized star hypergraph  consisting of $R$ ``inner" nodes and $N$ ``outer" nodes:   we take $R/N \rightarrow 0$ in the thermodynamic limit.  Allowed unitaries in the RUC act on a single outer vertex and two  inner vertices:  all such pairs are allowed.   See Figure  \ref{fig:3star}.

As on the  star graph under  2-local dynamics, we can understand 3-local dynamics on the generalized star by focusing on the fraction $s$ of inner vertices which are infected at any given time.   Denoting with $q$ the fraction of outer edges which are infected, the probability that $s$ transitions at each time step is  \begin{subequations}\begin{align}
\mathbb{P}\left(s\rightarrow s+\frac{2}{R}\right) &= (p_3+p_2) q(1-s)\left(1-s-\frac{1}{R}\right) , \\
\mathbb{P}\left(s\rightarrow s+\frac{1}{R}\right) &= 2(p_3 + p_2) s\left(1-s\right) + 2N(p_1+p_2) q(1-s)\left(1-s-\frac{1}{R}\right), \\
\mathbb{P}\left(s\rightarrow s-\frac{1}{R}\right)  &= 2p_1 s(1-s) + 2N(p_1+p_2)s\left(s-\frac{1}{R}\right), \\
\mathbb{P}\left(s\rightarrow s-\frac{2}{R}\right)  &= p_1 s\left(s-\frac{1}{R}\right).
\end{align}\end{subequations}
At early times $s\propto R^{-1}$ and $q\propto N^{-1}$.  Note that the rates of transition are the above probabilities multiplied by $N$ (the total number of vertices, in the thermodynamic limit).  The dominant infection-spreading at these early times are associated with unitaries that act on one infected inner node, and two uninfected nodes: one inner and one outer.   This means that $Rs$ undergoes a biased random walk from 0 to $R$.   Let us now ask -- given that at some time the random walk is at $Rs=m$, what is the probability $\mathfrak{p}_m$ that the random walk reaches $m=0$ before $m\sim R$?   This probability obeys the recursive equations $\mathfrak{p}_0=1$ and \cite{redner} \begin{equation}
\mathfrak{p}_m = \frac{p_3+p_2}{p_3+p_2+p_1} \mathfrak{p}_{m+1} + \frac{p_1}{p_3+p_2+p_1} \mathfrak{p}_{m-1}. 
\end{equation}
In the $R\rightarrow \infty$ limit, this set of equations is solved by \begin{equation}
\mathfrak{p}_m \approx \left(\frac{p_1}{p_2+p_3}\right)^m.
\end{equation}
What this means is that once a single vertex in the inner core is infected, there is a rather  high probability  that the inner core never becomes uninfected.   Generalizing this logic, we conclude that for any finite $m$, there is a finite probability that $Rs \ge m$ for all future times.   What this means is that we may approximate \begin{equation}
\frac{\mathrm{d}s}{\mathrm{d}t} \approx \frac{N}{R} (p_3+p_2-p_1)s,
\end{equation}
and so after an initial time \begin{equation}
t_{\mathrm{init}} \approx \frac{R\log R}{N(p_3+p_2-p_1)},
\end{equation}
a finite  fraction of the nodes  inner core are infected.   So long as $(R\log R) /N \rightarrow 0$ in the thermodynamic limit, $t_{\mathrm{init}} \rightarrow 0$.   

Once a finite fraction of the inner core is infected, then we may approximate \begin{equation}
\frac{\mathrm{d}q}{\mathrm{d}t} \approx (p_3 + 2p_2 + p_1)s_*(2-s_*)(1-q) - (2p_1+p_2) q,
\end{equation}
where \begin{equation}
s_* = \frac{p_3+p_2-p_1}{p_3+2p_2+p_1}
\end{equation}
is the fraction of inner vertices that are infected once $t\gg t_{\mathrm{init}}$.  The time  it takes for $q$ to be O(1) is  finite in thus given by a constant in the thermodynamic limit. By construction, we  have shown that there exist 3-local RUCs with small local Hilbert space dimensions, for which $\lambda_{\mathrm{L}} = \infty$.

\subsection{2-Local Entanglement Dynamics}
\label{app:dynamics}

Here we provide a more serious bound on the growth of entanglement for the random unitary circuit than the one found in the main text.   For simplicity, we focus on the 2-local case. The discussion below follows \cite{nahum16}, which studied the growth of entanglement on regular lattices.   As in the main text, we assume that our initial state is a product  state.   To bound the growth of entanglement we will need the following two inequalities for the von Neumann entropy \cite{chuang}: \begin{equation}
|S_A - S_B| \le S_{AB} \le S_A +S_B.
\end{equation}
Here $A$ and $B$ are disjoint subsets of the vertex set $V$.   This implies that \begin{equation}
S_A(t) \le S_{A+v}(t) + M\log 2, \;\;\;\;\;  S_A(t) \le S_{A-v}(t) + M\log 2.  \label{eq:2entropyineqs}
\end{equation}
Here $A+v=A\cup \lbrace v\rbrace$ and $A-v=A\cap \lbrace v\rbrace^{\mathrm{c}}$.   The second useful fact   that we will  need is that $S_A[|\Psi\rangle] = S_A[ U_{ij}|\Psi\rangle]$ if either $\lbrace i,j\rbrace \subset A$ or $\lbrace i,j\rbrace \cap A = \emptyset$:  this is easy to show by writing out the explicit formula for $S_A$ in terms of a partial trace.   

\begin{figure}
 \centering
 \includegraphics[width=3in]{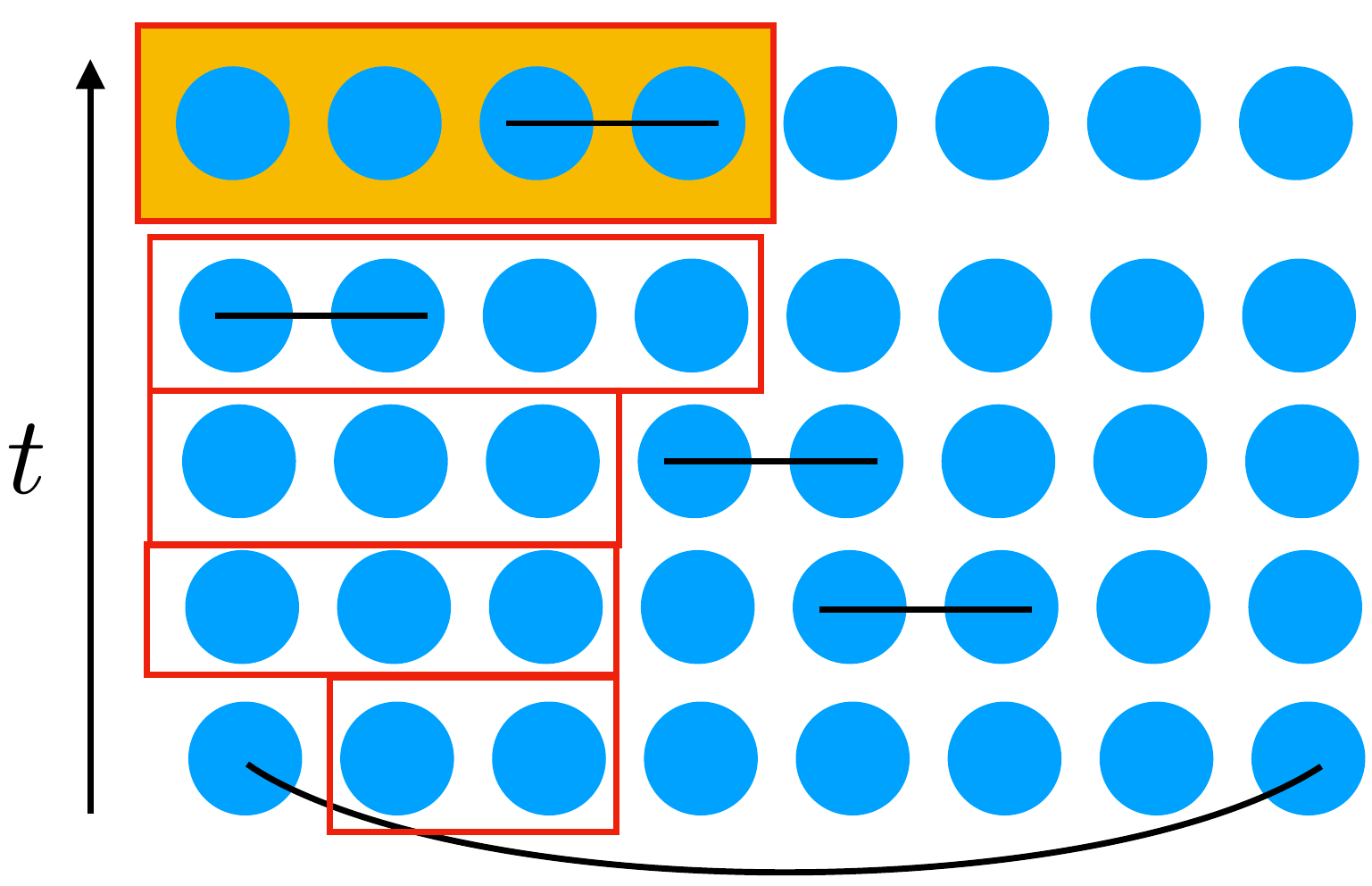}
 \caption{Bounding the entanglement generation in a 2-local RUC with $L=8$ vertices.  After 5 time steps, we see that the set $B(t)$ has changed during 2 time steps, and so we conclude that $S_A(t) \le 2M\log 2$.}
 \label{fig:cutting}
 \end{figure}

(\ref{eq:2entropyineqs}) may be used to get sharper bounds on the growth of entanglement.   In particular, the result can be  phrased as optimizing the following ``cutting" problem on a graph: choose a (possibly) time-dependent subset $B(t)\subseteq V$ of vertices for which at every discrete time $t$,  the unitary $U_{i_t j_t}(t)$ acts either entirely within $B(t)$ or entirely outside $B(t)$.  At time steps where the subset $B(t)$ changes, we either remove or add one vertex to $B(t)$.   Thus we conclude that \begin{equation}
S_B(t) \le  \left\lbrace  \begin{array}{ll} S_B(t-\frac{1}{L}) &\   B(t) = B(t-\frac{1}{L}) \\  S_B(t-\frac{1}{L}) + M\log 2 &\ B(t) \ne  B(t-\frac{1}{L}) \end{array}\right..  \label{eq:cutting}
\end{equation}
For simplicity, we simply write $S_B(t)$ instead of $S_{B(t)}(t)$.  We now look for fluctuating subsets $B(s)$ with the boundary condition $B(t)=A$, and arbitrary initial subset $B(0)$.   As the initial state is a tensor product, $S_B(0)=0$, and so we conclude from (\ref{eq:cutting}) that  \begin{equation}
S_A(t) = S_B(t) \le  M\log 2  \times n_{\mathrm{cut}}[B(t)]
\end{equation}
 where $n_{\mathrm{cut}}$ denotes the number of time steps at which the subset $B$ changes. See Figure  \ref{fig:cutting} for an illustration of the algorithm described above.  
 
  The entropy $S_A(t)$ is clearly bounded by the choice of fluctuating subset $B(t)$ with the minimal number of cuts.   In the limit of large $M$,  $S_A(t)$ was observed to be given \emph{exactly} by the minimal number of cuts (up to a factor of $M\log  2$) \cite{nahum16}.

\section{The Cheeger Inequality}
\label{app:graphs}
In this appendix we show that $\Lambda$ has a finite spectral gap on any graph with sufficient ``nonlocality".   The  result goes by the name of the Cheeger inequality:  the derivation below follows \cite{chung} and we review it here for completeness.

Let $\phi_v$ be any eigenvector of $\Lambda$, and let $\lambda>0$ be the associated non-zero eigenvalue:  \begin{equation}
\lambda \phi_v  = k_v\phi_v - \sum_{u:uv\in E} \phi_u.
\end{equation}
The spectral gap is given by the smallest possible choice of $\lambda$, and so we will simply look for bounds on $\lambda$.   Let $V_+\subset V$ be the vertices where $\phi_v > 0$.    Note that $V_+\ne V$ and $V_+\ne  \emptyset$  as $\Lambda$ is a symmetric matrix with orthogonal eigenvectors, and $\Lambda$ has a null vector $(1,\ldots,1)$.      We may write  \begin{equation}
\lambda = \dfrac{\displaystyle\sum_{v\in V_+}  \phi_v \left(k_v\hat{\phi}_v - \sum_{u:uv\in E} \phi_u\right) }{\displaystyle\sum_{v\in V_+} \phi_v^2}.
\end{equation}
Defining \begin{equation}
\hat{\phi}_v \equiv \left\lbrace \begin{array}{ll}  \phi_v &\  v\in V_+ \\ 0 &\ \text{otherwise} \end{array}\right.
\end{equation}
we obtain \begin{align}
\lambda &= \dfrac{\displaystyle\sum_{v}  \hat{\phi}_v \left(k_v\hat{\phi}_v - \sum_{u:uv\in E} \phi_u\right) }{\displaystyle\sum_{v} \hat{\phi}_v^2} > \dfrac{\displaystyle\sum_{v}  \hat{\phi}_v \left(k_v\hat{\phi}_v - \sum_{u:uv\in E} \hat{\phi}_u\right) }{\displaystyle\sum_{v} \hat{\phi}_v^2}  = \frac{\displaystyle\sum_{uv\in E}(\hat{\phi}_u-\hat{\phi}_v)^2} {\displaystyle\sum_{v} \hat{\phi}_v^2} \notag \\
&> \frac{\displaystyle\sum_{uv\in E}(\hat{\phi}_u-\hat{\phi}_v)^2} {\displaystyle\sum_{v} \hat{\phi}_v^2}\frac{\displaystyle\sum_{uv\in E}(\hat{\phi}_u+\hat{\phi}_v)^2} {2\displaystyle\sum_{v}k_v \hat{\phi}_v^2}
\end{align}
Now, using the identity (for real numbers) $a^2+b^2 \ge 2ab$, and thus \begin{equation}
(a-b)^2(c+d)^2 + (a+b)^2(c-d)^2 \ge 2\left|\left(a^2-b^2\right)\left(c^2-d^2\right)\right|,
\end{equation}
together with the Cauchy-Schwarz inequality on the vectors $\hat{\phi}_v$ and $\sqrt{D}\hat{\phi}_v$, we obtain  \begin{equation}
\lambda > \frac{1}{2} \left(\dfrac{\displaystyle \sum_{uv\in E} \left|\hat{\phi}_u^2-\hat{\phi}_v^2\right|}{\sum_v \sqrt{k}_v\hat{\phi}_v^2}\right)^2.
\end{equation}

Without loss of generality, label the vertices such that $\phi_1\ge \phi_2\ge\cdots\ge\phi_L$.   Then \begin{equation}
\sum_{uv\in E} \left|\hat{\phi}_u^2-\hat{\phi}_v^2\right| = \sum_{u=1}^L \sum_{v>u} A_{vu} \sum_{\ell=u}^{v-1} \left(\hat{\phi}^2_\ell - \hat{\phi}^2_{\ell+1}\right)
\end{equation}
Denote with $\mathcal{E}_i$ the number of  edges between $\lbrace 1,\ldots,i\rbrace$ and $\lbrace i+1,\ldots,L\rbrace$.  Then \begin{equation}
\sum_{uv\in E} \left|\hat{\phi}_u^2-\hat{\phi}_v^2\right|  = \sum_{\ell=1}^{L-1} \mathcal{E}_\ell \left(\hat{\phi}^2_\ell - \hat{\phi}^2_{\ell+1}\right)
\end{equation}
In the mathematics literature, it is more helpful to express $\mathcal{E}_i$ in  terms of a constant which we define as \begin{equation}
h_i \equiv \dfrac{\mathcal{E}_i}{\displaystyle \min \left( \sum_{v=1}^i k_v, \sum_{v=i+1}^L k_v\right)}.
\end{equation}
  $h_i$ denotes the fraction of edges that go between $\lbrace 1,\ldots, i \rbrace$ and $\lbrace i+1, \ldots, L\rbrace $.    In fact, the Cheeger constant, defined  as \begin{equation}
h_G = \min_{A\subset V} \frac{|E_{A,A^{\mathrm{c}}}|}{\min(|E_A|,|E_{A^{\mathrm{c}}}|)}
\end{equation}
where $E_{A,A^{\mathrm{c}}}\subset E$ consists  of the edges between $A$ and $A^{\mathrm{c}}$, is well known to be \emph{finite} in the thermodynamic limit for many families of random graphs \cite{chung}.   This is a quantitative measure of how ``locally treelike" the graph $G$ is.   Clearly,  $h_G \le h_i$.   So \begin{equation}
\sum_{uv\in E} \left|\hat{\phi}_u^2-\hat{\phi}_v^2\right|  \ge \sum_{\ell} h_G\left(\hat{\phi}^2_\ell - \hat{\phi}^2_{\ell+1}\right) \sum_{j\le \ell} k_j = h_G \sum_\ell k_\ell \hat{\phi}_\ell^2
\end{equation}and hence \begin{equation}
\lambda > \frac{h_G^2}{2}  \left(\frac{\displaystyle  \sum_\ell k_\ell \hat{\phi}_\ell^2}{\displaystyle   \sum_v \sqrt{k_v}\hat{\phi}_v^2}\right)^2 > \frac{h_G^2}{2} \frac{k_{\mathrm{min}}^2}{k_{\mathrm{max}}}.
\end{equation}
On any graph where $h_G>0$ and $k_{\mathrm{min}}/k_{\mathrm{max}}$ is a finite positive constant in the thermodynamic limit, this proves that the spectral gap of $\Lambda$ is strictly positive.

\end{appendix}

\bibliographystyle{unsrt}
\addcontentsline{toc}{section}{References}
\bibliography{sykgraphbib}

\end{document}